\documentclass[11pt]{article}

\usepackage{amsmath}
\usepackage{graphicx}
\usepackage{indentfirst}
\usepackage{amssymb}
\usepackage{cite}
\usepackage{color}
\usepackage{subfigure}
\usepackage{varwidth}

\begin{document}

\title{Holographic image features of an AdS black hole in  Einstein-power-Yang-Mills gravity  }

\date{}
\maketitle

\begin{center}
\author{Xin-Yun Hu,}$^{a}$\footnote{E-mail: huxinyun@126.com}
\author{Ke-Jian He,}$^{b}$\footnote{E-mail: kjhe94@163.com}
\author{Xiao-Xiong Zeng}$^{c, b}$\footnote{E-mail:  xxzengphysics@163.com (Corresponding author)}
\\

\vskip 0.25in
$^{a}$\it{College of Economic and Management, Chongqing Jiaotong University, Chongqing 400074, People's Republic of China}\\
$^{b}$\it{Department of Mechanics, Chongqing Jiaotong University, Chongqing 400074,  People's Republic of China}\\
$^{c}$\it{College of Physics and Electronic Engineering, Chongqing Normal University, Chongqing 401331, People's Republic of China}\\
\end{center}
\vskip 0.6in
{\abstract
{  By utilizing the AdS/CFT correspondence, we investigate the holographic image of an AdS black hole  in Einstein-power-Yang-Mills gravity. The AdS boundary hosts a Gaussian oscillation source, which induces a lensed response on the opposite side of the boundary during  propagation through bulk spacetime. The optical system assists observers at the north pole to continuously capture holographic images that show an axisymmetric bright ring known as the Einstein ring. As the observation position shifted, the bright ring gradually transformed into a  luminous arc and eventually transitioned into a light point. Simultaneously, we examine the impact of variations in relevant physical quantity on the ring, and present the corresponding brightness curve. The results indicate that as the temperature $T$ and nonlinear Yang Mills charge parameter $q$ increase, the ring radius also increases, while an increase in the chemical potential $u$ leads to a decrease. However, the peak brightness curve of the ring invariably decreases as the values of  $T$, $u$, and $q$ increase, albeit to varying degrees. Upon comparing the outcomes of geometric optics, it can be observed that the position of the ring in holography images is consistent with that of the photon ring.}}

\thispagestyle{empty}
\newpage
\setcounter{page}{1}\

\section{Introduction}
\label{sec:intro}
General Relativity (GR) is widely acknowledged as the preeminent theory of gravitation due to its extensive validation through a multitude of observational studies and experiments. The predictions of GR suggest that black holes, immensely compact celestial objects in our universe, have garnered significant attention from physicists. Recently, the Event Horizon Telescope Collaboration (EHT) has revealed unprecedented images of the black hole situated at the core of radio galaxy M87$^*$\cite{EventHorizonTelescope:2019dse,EventHorizonTelescope:2019uob,EventHorizonTelescope:2019jan,EventHorizonTelescope:2019ths,EventHorizonTelescope:2019pgp,EventHorizonTelescope:2019ggy} and the Milky Way's center Sagittarius A$^*$ (Sgr A$^*$)\cite{EventHorizonTelescope:2022wkp,EventHorizonTelescope:2022vjs,EventHorizonTelescope:2022wok,EventHorizonTelescope:2022exc,EventHorizonTelescope:2022urf,EventHorizonTelescope:2022xqj}, providing  compelling evidence for the existence of black holes.  One of the most remarkable characteristics exhibited by these images is the presence of a dark region within a luminous ring structure. According to the GR,  the gravitational field in the vicinity of a supermassive black hole is so intense that it causes a significant curvature of spacetime. As a result, the photon attempting to escape this region are ultimately captured by the black hole. Hence,  from the observer's perspective, a dark area will be seen, which is called the black hole shadow\cite{Synge:1966okc,Bar,Chandrasekhar,Cunha:2018acu}. On the other hand, due to the powerful gravitational force around a black hole, a mass of matter will be accreted together to form what is called an accretion disk. During the accretion process, part of the gravitational potential energy is converted into thermal radiation. Therefore, the accretion disk is generally regarded as a bright light source near the black hole. A portion of the light emitted from the accretion disk reaches the observer, resulting in the perception of a luminous structure, commonly referred to as the photon ring. Based on the findings reported by the EHT, the angular size of the photon ring surrounding Sgr A$^*$  aligns closely with the shadow critical curve radius predicted by the GR, with a deviation of no more than  10$\%$\cite{EventHorizonTelescope:2022wkp,EventHorizonTelescope:2022vjs,EventHorizonTelescope:2022wok,EventHorizonTelescope:2022exc,EventHorizonTelescope:2022urf,EventHorizonTelescope:2022xqj}. With the enhancement of EHT resolution and the emergence of experimental advancements such as the image of the black hole at the center of the Milky Way, it is inevitable that these developments will impose higher demands on the theoretical study of black hole imaging. In the theoretical investigation of black hole imagery, photon rings and black hole shadows have garnered significant attention. This is because their study not only elucidates the fundamental properties of black holes but also serves as a robust tool for constraining various gravitational models \cite{Gralla:2019xty,Wei:2020ght,Zeng:2021dlj,Zeng:2021mok,Li:2021riw,Zeng:2022pvb,He:2021htq,He:2022yse,Guo:2021bwr,He:2024yeg,Guo:2023grt,He:2024qka,Yan:2021ygy,Zhong:2021mty,Wei:2013kza,He:2025rjq,Tsukamoto:2014tja,He:2024amh,Hou:2022eev,Hu:2020usx,He:2022aox}.

Despite extensive research on the properties of black hole  shadows and their associated dynamics, which has provided substantial insights into black hole physics, many mysteries remain to be unraveled using more realistic models.
It is well known that the ($D + 1$)-dimensional gravity theory in Anti-de Sitter (AdS) can be associated with the $D$-dimensional Conformal field theory (CFT) due to the AdS/CFT correspondence  (Anti-de Sitter spacetime/Conformal field theory), which is  also known as holography theory \cite{Maldacena:1997re,Gubser:1998bc,Witten:1998qj}. As a pioneering endeavor, Hashimoto et al. constructed  holographic images of the Schwarzschild AdS black hole in the bulk using  a given response function of Quantum field theory (QFT) on the boundary within the framework of AdS/CFT duality\cite{Hashimoto:2019jmw,Hashimoto:2018okj}. The results indicate the presence of a distinct circular structure in two-dimensional spherical thermal holographic images, namely,  the Einstein ring, and the size of the ring is consistent with the size of the black hole photon ring observed by geometric optics. Considering that  real quantum materials are typically engineered with a finite chemical potential\cite{Liu:2022cev}, Liu et al. investigated the  holographic image of a charged AdS black hole, where the complex scalar field serves as a probing wave in the bulk spacetime. They observed that the radius of ring  remained unaffected by changes in the chemical potential, while it exhibited a distinct dependence on the temperature of the boundary system.
Then, the study of a black hole that is dual to a superconductor was also conducted in\cite{Kaku:2021xqp}, where the electric current of the superconductor under  localized time-periodic external electromagnetic field was considered. They also acquired images that depict the discontinuous changes in the size of the photon ring. Moreover, the investigation into holographic imaging of black holes has been expanded to encompass various gravitational frameworks \cite{Aslam:2024bmx,Hu:2023mai,Hu:2023eow,Zeng:2023zlf,Zeng:2023ihy,Zeng:2023tjb,Hu:2023eoa,Li:2024mdd,He:2024bll,Zeng:2024ptv,He:2024mal,Gui:2025see}. In these studies, the characteristics of Einstein ring structures across various parameter spaces have been investigated. It has been discovered that, for a given set of wave sources and optical systems, holographic images serve as an effective tool to differentiate between distinct types of black holes.

In recent years, there has been a significant surge of interest in gravitational theories that incorporate non-linearity in the Maxwell fields. The discovery of the nonsingular nature of black hole solutions in nonlinear electrodynamics  opens up a new perspective on the physics of black holes\cite{Ayon-Beato:1998hmi}. Among them, the class of black hole solutions in power Maxwell invariant  theory is derived,  where the Lagrangian density is given by $({F}_{\mu\nu} {F}^{\mu\nu})^q$ and $q$ represents an arbitrary rational number\cite{Maeda:2008ha}. Considering the Yang-Mills (YM) fields as an exception to the black hole  no-hair theorem in both GR \cite{Bizon:1990sr} and conformal gravity \cite{Riegert:1984zz,Fan:2014ixa}, more investigations have been conducted on  nonlinear models  involving  the coupling of nonabelian YM fields  with gravity in the realm of GR\cite{Mazharimousavi:2009mb}. As mentioned in \cite{Mazharimousavi:2009mb}, there exists a black hole solution  sourced by the power of YM invariant, i.e., $({F}^{(a)}_{\mu\nu} {F}^{(a)}{\mu\nu})^q$, where ${F}^{(a)}_{\mu\nu}$  represents the YM field with its internal index ranging in  $1\leq a \leq \frac{(D-1)(D-2)}{2}$. It can be observed that such solutions successfully recover the $D$-dimensional Einstein-Yang-Mills (EYM) black holes in AdS spacetime for the case of  power exponent is taken as $q=1$ \cite{HabibMazharimousavi:2007fst,Mazharimousavi:2008ap}. The study of black hole properties within the framework of Einstein-power-Yang-Mills (EPYM) gravity has revealed novel and intriguing phenomena, including  van der Waals phase transitions and critical behavior\cite{Zhang:2014eap,Hendi:2016usw},  Joule-Thomson expansion\cite{Biswas:2021uop,Du:2023kgj}, as well as  quasinormal mode  and optical properties of the black hole\cite{Gogoi:2023ffh}.

Motivated by the aforementioned research, it is justified to construct holographic images of a black hole within the framework of EPYM gravity. Fundamentally, the photon ring exhibits variations contingent upon the specific bulk dual geometry, thereby inducing alterations of  corresponding holographic response. Therefore, we are tempted to investigate the behavior of the lensed response for the black hole in EPYM  gravity and construct holographic images, so as to study  the potential impacts  of the relevant physical quantities on the resulting holographic images. As an essential reference, the alteration of the power exponent $q$ will evidently impact the geometric structure of spacetime. Our concern lies in whether this alteration will be applied to the holographic image and how it will  change the features of images.  On the other hand, in the background of EPYM gravity,  there is a modification to the chemical potential of the boundary system.  We delve deeper into investigating whether the chemical potential has distinct implications for holographic images when compared to those arising from the charged black hole in \cite{Liu:2022cev}.  This could potentially serve as an effective means to differentiate between these two gravitational models.

The remainder of the present paper is outlined as follows. In Section 2, we present a holographic construction of the Einstein ring for an AdS black hole, where the scalar field serves as a probing wave in the bulk spacetime.   In Section 3,  we provide a holographic setup of an AdS black hole under EPYM gravity and extract  its corresponding lensed response function for analysis. In Section 4, we investigate the formation of Einstein rings within our theoretical analysis by using an optical system composed of a convex lens and spherical screen. Furthermore, we  not only present the brightness curve of the resulting ring but also conduct a comparative analysis with the ring obtained through optical approximation. The last section is devoted to the conclusion.

\section{Construction of holographic images }
Due to the phenomenon of gravitational lensing, the emitted light from a distant galaxy can undergo distortion and manifest as an arc or multiple separate images when it passes through an intermediate galaxy along the line of sight. In cases where two galaxies are perfectly aligned, the distorted light forms a ring-shaped structure that surrounds the foreground galaxy, commonly known as an Einstein ring. The  Einstein ring serves as one of the most compelling manifestations of GR in the cosmos, providing a unique opportunity to investigate colossal galaxies. If the intermediate galaxy were to be substituted with a black hole, the resulting observed image would inherently yield an abundance of valuable insights regarding the characteristics and properties of the black hole. From this perspective, Hashimoto et al. constructed a holographic Einstein ring of the black hole in Schwarzchild-AdS$_4$ spacetime(Sch-AdS$_4$) based on the AdS/CFT correspondence. In  \cite{Hashimoto:2019jmw,Hashimoto:2018okj}, the QFT is considered on $R_t \times S^2$ and  an external source  is introduced on $S^2$, where the external source is mapped to the boundary condition of the bulk field. The probe wave is emitted from the source located on the AdS boundary, diffracted by the black hole as it traverses through the spacetime of the black hole, and ultimately reaches another point on the AdS boundary, resulting in a lensed response. The asymptotic data of the bulk field corresponds to the response function in relation to an external source, and based on this response function, holographic images of an AdS black hole can be constructed. Consequently, they discovered a correspondence  between the resultant Einstein ring and the size of the black hole photon ring determined through geometric optics calculations.

\begin{figure}[h]
\centering 
\includegraphics[width=0.50\textwidth]{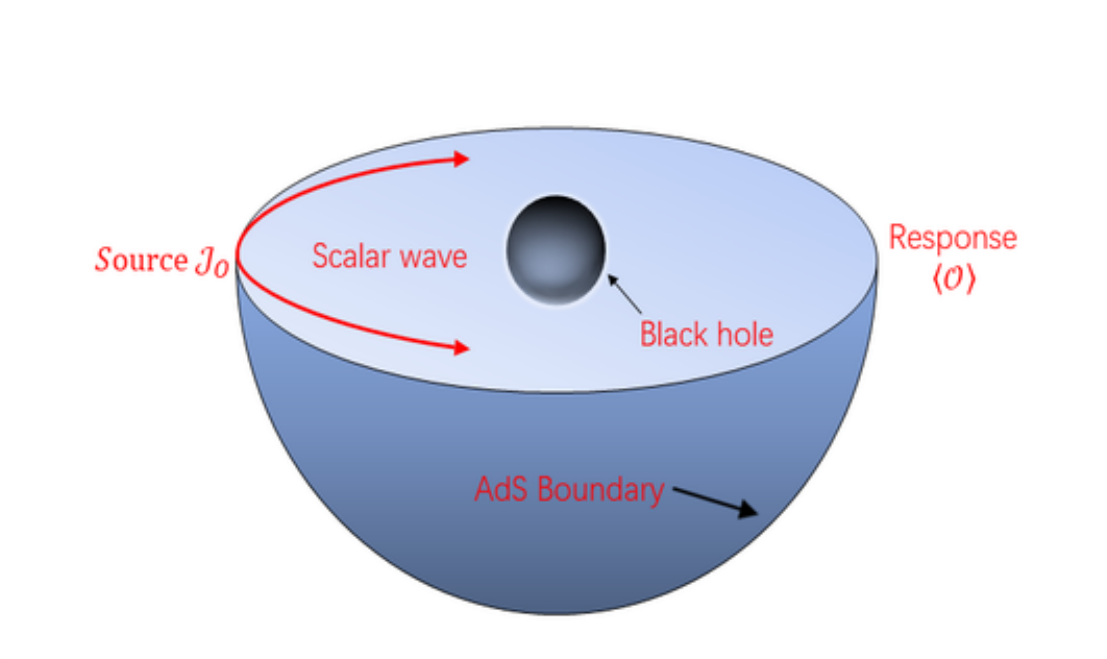}
\caption{\label{fig1}  A monochromatic Gaussian source is located at a point on the AdS boundary, and its response is observed at another point on the same boundary.  }
\end{figure}

In this study, we employ this concept to the (2 + 1)-dimensional boundary CFT on a 2-sphere $S^2$ at the finite temperature and chemical potential. We consider placing a time-periodic localized Gaussian source with frequency $\hat{\omega}$ on one side of the AdS boundary, as illustrated in Fig.1. The (2+1)-dimensional boundary CFT on the 2-sphere $S^2$ is dual to a black hole in the global AdS$_4$ spacetime,  or a massless bulk scaler field in this spacetime.  Motivated by the boundary conditions imposed by the time period, the scalar wave generated by the source can be injected into the bulk from the AdS boundary and propagated within the spacetime of a black hole in the bulk. The selected Gaussian source, when chosen as an axially symmetric monochromatic oscillating Gaussian source and localized at the south pole of the AdS boundary $\theta = \pi$, can be expressed as
\begin{align}
{J}_{\mathcal{O}}(\chi_e,\theta)=e^{-i \hat{\omega}  \chi_e} \mathcal{H}(\theta), \label{J1}
\end{align}
where
\begin{align}
\mathcal{H}(\theta)= \frac{1}{2 \pi \sigma^2} \text{exp} \left[-\frac{(\pi-\theta)^2}{2 \sigma^2}\right]. \label{J2}
\end{align}
In which, $\chi_e$ and $\theta$ represent coordinates,  $\sigma$ is the width of  the scalar wave. In the  case of $\sigma \ll 1$, the negligible value of the Gaussian tail allows for the decomposition of the Gaussian function into scalar spherical harmonics, that is
\begin{align}\label{Sph1}
 \mathcal{H} (\theta)\simeq \sum _{\zeta=0}^{\infty }C_{\zeta0}Y_{\zeta0}(\theta).
\end{align}
The corresponding coefficients  $C_{\zeta0}$ of the spherical harmonics $Y_{\zeta0}$ is
\begin{align}\label{Sph2}
C_{\zeta0} \equiv (-1)^\zeta \sqrt{\frac{\zeta+\frac{1}{2}}{2 \pi }} \text{exp} \left[-\frac{1}{2} (\zeta+\frac{1}{2})^2 \sigma^2\right],
\end{align}
where  $\zeta$ denotes the magnetic quantum number.
\begin{figure}[h]
\centering 
{\includegraphics[width=.8\textwidth]{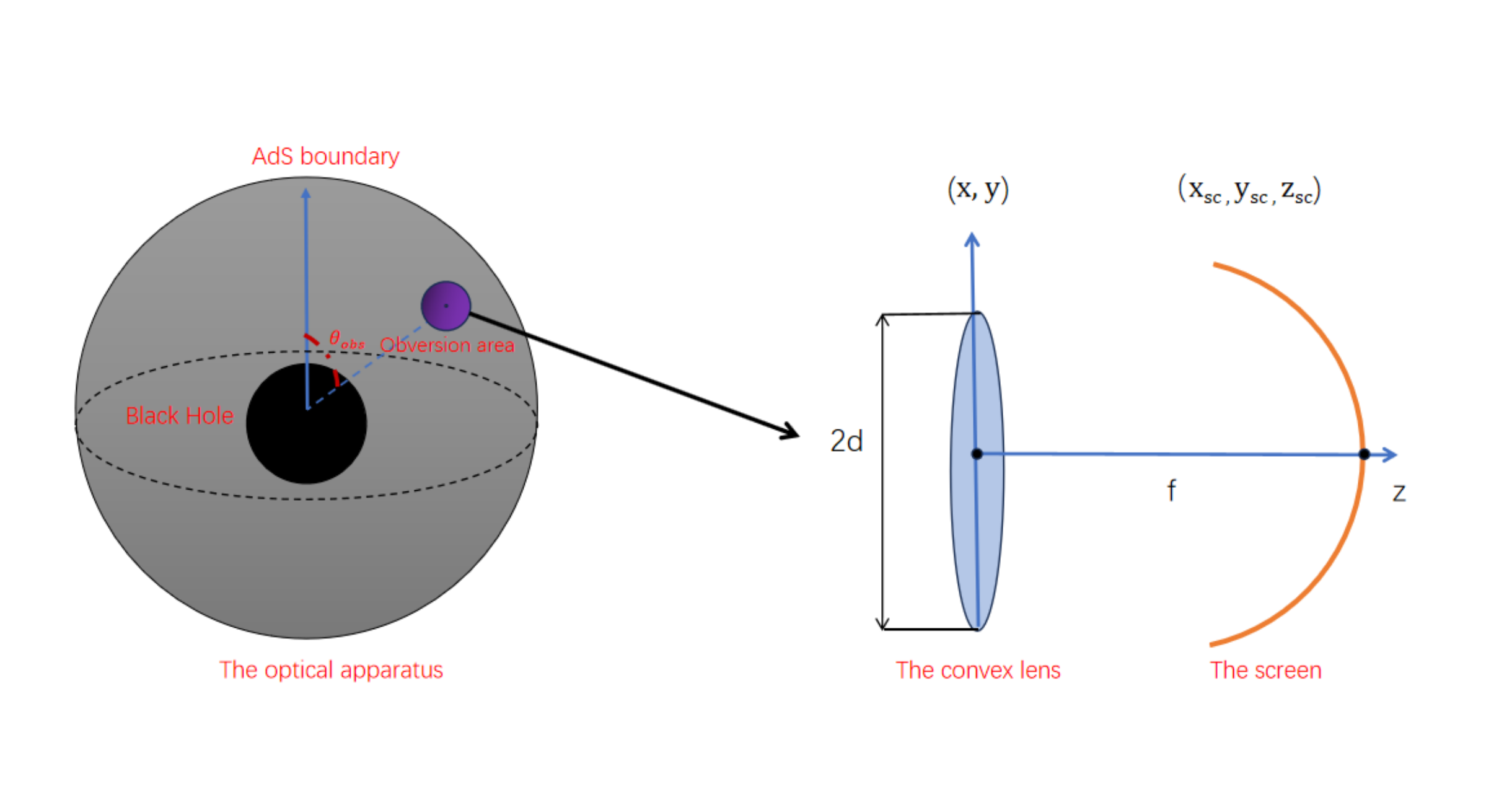}}
\caption{\label{fig2}  The schematic diagram illustrates the construction of the black hole image using the response function, with the observation area indicated by the purple region on the AdS boundary.}
\end{figure}

The probe wave undergoes diffraction by the black hole as it propagates within the bulk, resulting in its arrival at other points on the AdS boundary and leading to the corresponding response function $\langle \mathcal{O} (x)\rangle$.  The lensed response is crucial as it encapsulates valuable information regarding the spacetime geometry within the bulk. However, obtaining the response function on the AdS boundary is  just one of the prerequisites; achieving imaging of the dual black hole also requires a virtual optical system in three-dimensional flat space, which consists of a convex lens and a spherical screen, see the right side of Fig.2. The response function within the local range on the boundary, specifically indicated by the purple circle in the left part of Fig.2, is extracted and subsequently examined using the imaging system.  During the process of transforming the response function $\langle \mathcal{O} (x)\rangle$ into an image of a black hole on the screen $\Psi _{sc} (\hat{x}_{sc})$, satisfying
\cite{Hashimoto:2019jmw,Hashimoto:2018okj}
\begin{align}
\Psi _{sc} (\hat{x}_{sc})=\int _{\left| \hat{x}\right| <d} d^2 x \langle \mathcal{O} (\hat{x})\rangle e^{-\frac{i \hat{\omega} }{f} \hat{x} \cdot \hat x_{sc}}.\label{TR}
\end{align}
where $\hat{x}=(x,y,z)$ is the Cartesian-like coordinate on the boundary $S^2$, while $\hat x_{sc}=(x_{sc}, y_{sc}, z_{sc})$ is on the screen.  Actually, Eq.(\ref{TR}) corresponds to the Fourier transform of the response function within a specific local range.  In the particular imaging system, the convex lens can be regarded as a converter that effectively transforms the plane wave into spherical wave and projects it onto a curved screen. The observer is positioned at the central point of the observation area, denoted as $(\theta, \varphi)= (\theta_{obs},0)$. Subsequently, we adjust the position of convex lens  on the three-dimensional flat space to achieve $(x_{cl},y_{cl})=(x, y, 0)$. The response function obtained from the observation area is projected as a planar wave onto a thin lens (referred to as the incident wave $\Psi_{iw} (\hat{x})$), which is then transformed by the thin lens into a spherical wave (referred to as the outgoing wave $\Psi_{ow} (\hat{x})$) and converges at the focal  located at $z = f$.  According to the wave optics, the transformation relationship is
\begin{align}
\Psi_{ow} (\hat{x})=e^{-i {\hat{\omega} }\frac{|\hat x|^2}{2f}}\Psi_{iw} (\hat{x}), \label{TX2}
\end{align}
In principle, what we require is an ultra-thin convex lens with a radius $d$ and focal length $f$ that satisfy the condition of $f \gg d$. Then, the spherical screen is located at $(x,y,z)=(x_{sc},y_{sc},z_{sc})$, where ${x_{sc}}^2+{y_{sc}}^2+{z_{sc}}^2=f^2$. The wave equation after projecting the outgoing wave $\Psi_{ow} (\hat{x})$ onto the screen is recorded as
\begin{align}
\Psi_{sc}(\hat x_{sc})=\int _{\left| \hat{x}\right| \leq d} d^2 x \Psi_{ow} (\hat{x}) e^{-i {\hat{\omega} } \mathcal{N} }, \label{TS2}
\end{align}
where $\mathcal{N}$ is the distance from the lens point $(x, y, 0)$  to the screen point $(x_{sc}, y_{sc}, z_{sc})$. By combining Eq.(\ref{TX2}) and Eq.(\ref{TS2}), one can obtain that
\begin{align}
\Psi_{sc}(\hat x_{sc})=\int _{\left| \hat{x}\right| \leq d} d^2 x \Psi_{ow} (\hat{x}) e^{-i {\hat{\omega}} \mathcal{N} } \propto \int _{\left| \hat{x}\right| \leq d} d^2 x \Psi_{iw} (\hat{x}) e^{-i \frac{\hat{\omega} }{f} \hat x \cdot \hat x_{sc}}= \int _{\left| \hat{x}\right| \leq d} d^2 x \Psi_{iw} (\hat{x})\mathcal{W}(\hat{x})  e^{-i \frac{\hat{\omega} }{f} \hat x \cdot \hat x_{sc}}, \label{WF}
\end{align}
in which $\mathcal{W}(\hat{x})$  stands for the window function, that is
\begin{align}
   \mathcal{W}(\hat{x}) \equiv  \begin{cases}1, &  0<\left| \hat{x}\right| \leq d,  \\
    0, &\left| \hat{x}\right| > d. \label{WQ}
    \end{cases}
\end{align}
As demonstrated by Eq.(\ref{WF}), the wave observed on the screen can be associated with the incident wave through the application of Fourier transform within a finite domain, i.e.,the range of lens size.
In essence, there is a direct correlation between the image of the black hole on the screen and the response function.

\section{  Construction  the  holographic image of  an AdS black hole}

\subsection{ The AdS black hole solution in Einstein-power-Yang-Mills  gravity }
Initially, we commence with the $D$-dimensional action for EPYM gravity that incorporates a non-vanishing cosmological constant. Such action is given by\cite{Mazharimousavi:2009mb}
\begin{align}
I=\frac{1}{2}\int d^D x \sqrt{-g}(\mathcal{R}-\frac{(D-2)(D-1)\Lambda}{3}-\mathcal{F}^q),\label{action1}
\end{align}
in which $\mathcal{R}$ represents the Ricci Scalar, $q$ is a  real parameter that introduces non-linearities. The term of $\mathcal{F}$ denotes the  YM invariant,  written as
\begin{align}
\mathcal{F}=\mathbf{T_r}(F^{(a)}_{\lambda \eta} F^{(a)\lambda \eta}),\label{YM1}
\end{align}
and
\begin{align}
\mathbf{T_r}(\cdot)=\sum ^{(D-2)(D-1)/2}_{a=1}(\cdot).\label{YM2}
\end{align}
And, the YM field is \cite{Biswas:2022qyl}
\begin{align}
\mathcal{F}^{(a)}_{\lambda \eta}=\partial_\mu A^{(a)}_\nu -\partial_\nu A^{(a)}_\mu +\frac{1}{2\eta} C^{(a)}_{(b)(c)} A^{(b)}_\mu A^{(c)}_\mu,\label{YMfield}
\end{align}
where $C^{(a)}_{(b)(c)}$ stands for the structure constants of $[\frac{(D-2)(D-1)}{2}]-$ parameter Lie group $G$. Furthermore, $\eta$ represents  an arbitrary coupling constant, and $A^{(a)}_\mu$ denotes the $SO(D-1)$  gauge group YM potentials, which are described by the Wu-Yang ansatz \cite{Balakin:2015gpq,Balakin:2007nw}. The field equations are obtained by varying the action with respect to the spacetime metric $g_{\mu\nu}$, which are
\begin{align}
{G^{\mu }}_{\nu }+\frac{(D-2)(D-1)}{6}\Lambda {\delta^{\mu }}_{\nu }={T^{\mu }}_{\nu }, \label{Field
1}
\end{align}
and
\begin{align}
{T^{\mu }}_{\nu }=-\frac{1}{2}({\delta^{\mu }}_\nu \mathcal{F}^q -4 q \mathbf{T_r} (F^{(a)}_{\nu \lambda} F^{(a) \mu\lambda})\mathcal{F}^{q-1}),\label{Field2}
\end{align}
where $G_{\mu\nu}$ is the Einstein tensor. The variation of the gauge potentials $A^{(a)}$ leads to the YM  equation, expressed as
\begin{align}
\mathbf{d} (^\star \mathbf{F}^{(a)} \mathcal{F}^{q-1})+\frac{1}{\eta} C^{(a)}_{(b)(c)} \mathcal{F}^{q-1} \mathbf{A}^{(b)} \wedge {^\star} \mathbf{F}^{(c)}=0. \label{YME}
\end{align}
Here, $\mathbf{F}^{(a)}=\frac{1}{2}F^{(a)}_{\mu \nu}d x^\mu  \wedge d x^\nu $, $\mathbf{A}^{(b)}=A^{(b)}_\mu \wedge d x^\nu $, and $\star$ stands for the duality. The metric ansatz for $D$-dimensional spherically symmetric line element is chosen as
\begin{align}
ds^2=-N (r) d t^2+\frac{1}{N (r)} d r^2 +r^2 d \Omega_{D-2}^2,\label{metric1}
\end{align}
where $d \Omega_{D-2}^2$ is the line element associated with unit $D-2$ sphere. Depending on the dimensions $D$ and values of $q$, various black hole solutions can arise from the action given in Eq. (\ref{action1}).
Considering the condition that the  power exponent $ q \neq \frac{(D-1)}{4}$, the metric function $N(r)$ of $D$-dimensional EPYM black hole with  cosmological constant can be  provided by \cite{Biswas:2022qyl}
\begin{align}
N (r)=1-\frac{2 M} {r^{D-3}}-\frac{\Lambda r^2}{3}+\frac{[(D-3)(D-2)Q_{YM}^2]^q}{(D-2)(4 q-D+1)r^{4 q-2}},\label{Function1}
\end{align}
The parameter $M$ represents an integral constant that corresponds to the mass of the black hole, and $Q_{YM}$  denotes  a charge parameter which is linked to YM fields. In the AdS spacetime,  the  cosmological constant denoted as
\begin{align}
\Lambda=-\frac{(D-1)(D-2)}{2l^2},\label{CC}
\end{align}
where $l$ is  the radius of AdS spacetime. The focus of our study lies in the four-dimensional spacetime, i.e., $D=4$,  the metric function in Eq. (\ref{Function1}) can be expressed as follows,
\begin{align}
N (r)=1-\frac{2 M} {r}+\frac{ r^2}{l^2}+\frac{(2 Q_{YM}^2)^q}{2(4q-3)r^{4q-2}}.\label{Function2}
\end{align}
It is imperative for the power parameter $q$ to possess a positive value $q>0$ in order to ensure the fulfillment of the Weak Energy Condition (WEC) of the PYM term \cite{Mazharimousavi:2009mb}. And, the bound limit range of $q$ is constrained to $\frac{3}{4} < q <\frac{3}{2}$, thereby satisfying certain energy conditions, including the causality condition. In case of $q=1$, the aforementioned black hole solutions can be simplified into Einstein-Yang-Mills black holes\cite{HabibMazharimousavi:2007fst,Mazharimousavi:2008ap}. In this work, we consider two regions for the value of $q$, namely $\frac{3}{4} < q <1$ and $1< q <\frac{3}{2}$, in order to investigate whether other relevant physical quantities have the same impact on the holographic image under these two values of $q$. The event horizon of the black hole is located at $r_+$ , which corresponds to the largest positive real root of $N (r)=0$, and we can derive the expression of the mass of the black hole in terms of the event horizon radius,  that is
\begin{align}
M=\frac{1}{2}r_+ \left(1+\frac{r^2_+}{l^2}+\frac{2^{q-1}Q_{YM}^{2 q}}{(4 q-3)r^{4 q-2}_+}\right) .\label{MASS}
\end{align}

\subsection{The holographic setup of an AdS black hole in Einstein-power-Yang-Mills  gravity}
Now, we are embarking on constructing a holographic model for  an AdS black hole in EPYM gravity.  To introduce an affine parameter $\xi$, such that $\xi=1/r$, thereby we can get $N(r)=\frac{1}{\xi^2}N(\xi)$,  and the metric Eq. (\ref{metric1}) in   the new coordinate ($t,\xi,\theta,\varphi$) is substituted  with
\begin{align}
ds^2=\frac{1}{\xi^2}\left[-N (\xi) d t^2+\frac{1}{N(\xi)}d \xi ^2+ d \Omega^2\right].\label{metric3}
\end{align}
In this equation, $d \Omega^2=d \theta^2+\sin \theta^2 d \varphi^2$, and $N(\xi)$ is expressed as
\begin{align}
N(\xi)=1+\frac{1}{l^2 \xi ^2}-\frac{\xi }{l^2 \xi _+^3}-\frac{\xi }{\xi _+}+
\frac{2^{q-1 } \left[ \xi  \left(\frac{1}{\xi _+}\right){}^{3-4 q } \left(Q_{\text{YM}}^2\right){}^{q }- \left(\frac{1}{\xi }\right)^{2-4 q } \left(Q_{\text{YM}}^2\right){}^{q }\right]}{3-4 q},\label{metric4}
\end{align}
where $\xi_+=\frac{1}{r _+}$ denotes the location of black hole event horizon. The temperature of the boundary system is given by the Hawking temperature, which is
\begin{align}
T=\frac{\xi_+}{4 \pi }+ \frac{3}{4  \pi \xi_+ l^2}- \frac{2^{q-1 (Q^2_{YM})^q}}{4 \pi}\xi_+^{4q-1}.\label{Temperature}
\end{align}
The complex scalar field $\Phi$ is considered as probe field within the system, and its dynamics can be effectively described by the Klein-Gordon equation\cite{Liu:2022cev}, that is
\begin{align}
\mathcal{D}_\kappa \mathcal{D}^\kappa \Phi -\mathbb{M}^2 \Phi =0,\label{KG}
\end{align}
in which $\mathcal{D}_\kappa=\nabla_\kappa - i e A_\kappa $, and $e$  represents the electric charge  of the complex scalar field $\Phi$. The utilization of Eddington coordinates enables a more efficient resolution of the Klein-Gordon equation, that is
\begin{align}
\chi_e \equiv t+\xi_\ast=t-\int \frac{1}{N (\xi)} d\xi.\label{EC}
\end{align}
Then, the metric equation is rewritten  in smooth form as
\begin{align}
ds^2=\frac{1}{\xi^2}\left[-N (\xi) d \chi_e^2-2 d\chi_e d\xi+ d \Omega^2\right].\label{metric5}
\end{align}
For the sake of clarity, we will set $\mathbb{M}=-2$ and $e=1$ in the subsequent analysis. The asymptotic behavior of  the complex scalar field near the AdS boundary can be defined as
\begin{align}
&\Phi (\chi_e, \xi, \theta, \varphi)=\mathcal{J}_{\mathcal{O}} (\chi_e, \theta, \varphi)\xi_+ +
\langle \mathcal{O} \rangle \xi_+^2  + O (\xi^3).\label{SFAB}
\end{align}
The terms  $\mathcal{J}_{\mathcal{O}}$ and $\langle \mathcal{O} \rangle$ are the independent functions of  boundary coordinates $(\chi_e, \theta,\varphi)$  in accordance with the AdS/CFT dictionary. In addition,  $\mathcal{J}_O$ is regarded as the primary source for the boundary field theory and  its  corresponding expectation value of the dual operator, i.e., the response function, is determined by
\begin{align}
\langle \mathcal{O} \rangle _{\mathcal{J}_{\mathcal{O}}}=\langle \mathcal{O} \rangle-(\partial_{\chi_e}-i u) \mathcal{J}_{\mathcal{O}}. \label{RES}
\end{align}
Here, the chemical potential $u$ of the boundary system is  given by  $u= e \frac{2^{q -1} q  \left(Q_{\text{YM}}^2\right){}^{q }}{(4 q -3) Q_{\text{YM}}} \xi _+^{4 q -3}$ under holography. Additionally, the term $\langle \mathcal{O} \rangle$ represents the expectation value of the dual operator when the source is turned off. In  the symmetry of the spacetime background,  considering the source provided by Eq. (\ref{J1}), we can obtain the bulk solution, denoted as
\begin{align}
\Phi (\chi_e, \xi, \theta,\varphi)= e^{-i \hat{\omega}  \chi_e} \sum _{\zeta=0}^{\infty }C_{\zeta 0} \xi Z_\zeta(\xi) Y_{\zeta 0}(\theta). \label{BS}
\end{align}
The term $Z_\zeta$ satisfies the equation of motion, that is
\begin{align}
\xi^2 N(\xi)Z_\zeta''+ \xi^2  \left[N'(\xi)+2 i (\hat{\omega}- e A) \right]Z_\zeta' +  \left[(2-2 N(\xi))+\xi N'(\xi)-\xi^2 \zeta(\zeta+1)- i e \xi^2 A'\right]Z_\zeta=0, \label{ZMO}
\end{align}
in which $N'(\xi)= \partial_\xi N(\xi)$. Near the AdS boundary, the asymptotic behaviour of the $Z_\zeta$ is expressed as
\begin{align}
Z_\zeta=1+ \langle \mathcal{O} \rangle_{\zeta} \xi + O (\xi^2).  \label{ZAS}
\end{align}
Then, the relevant response function ${\langle \mathcal{O} \rangle} _{\mathcal{J}_{\mathcal{O}}}$ is derived, which is
\begin{align}
{\langle \mathcal{O} \rangle} _{\mathcal{J}_{\mathcal{O}}}= e^{-i {\hat{\omega} } \chi_e} \sum _{\zeta=0}^{\infty }C_{\zeta 0} \langle \mathcal{O} \rangle_{ \mathcal{J}_{\mathcal{O}} \zeta} Y_{\zeta 0}(\theta),\label{REF2}
\end{align}
where
\begin{align}
\langle \mathcal{O} \rangle_{ \mathcal{J}_{\mathcal{O}} \zeta}= \langle \mathcal{O} \rangle_\zeta+i \overline{\omega},\label{REF3}
\end{align}
and  $\overline{\omega}$ is represented as $\overline{\omega}=\hat{\omega}+u$. The extraction of the $\langle \mathcal{O} \rangle_{\zeta}$, which corresponds to solving Eq.(\ref{ZMO}),  is crucial for obtaining the  total response function  ${\langle \mathcal{O} \rangle} _{\mathcal{J}_{\mathcal{O}}}$. The radial equation Eq.(\ref{ZMO}) can be effectively solved by considering two crucial boundary conditions, $\mathbf{i})$ the AdS boundary $Z_\zeta(0)=1$; $\mathbf{ii})$ the regular boundary conditions imposed on the event horizon of the black hole $\xi=\xi_+$. Based on this premise, we use the pseudo-spectral method to obtain the numerical solution of $ Z_\zeta$ and derive $\langle \mathcal{O} \rangle_{\zeta}$ \cite{Hashimoto:2019jmw,Hashimoto:2018okj}. Taking into account the parameters associated with source and spacetime structure, specifically the frequency of  Gaussian source $\hat{\omega}$ and the power exponent $q$, the  typical profile of the total response function ${\langle \mathcal{O} \rangle} _{\mathcal{J}_O}$ is depicted in Fig.3 and 4. It has been demonstrated that the interference pattern indeed arises from the diffraction of the probe scalar field by the black hole in the bulk spacetime. In Fig.3, it can be observed that the amplitude of the total response function exhibits an increasing trend with the increasing values of $q$, while keeping other involved parameters at specific values such as $Q_{YM}=0.5, \xi_+=0.5$ and $\hat{\omega}=80$. When other parameters are fixed at $Q_{YM} = 0.5,\xi_+= 0.5$, and $q = 7/8$, an increase in frequency $\hat{\omega}$ leads to a significant reduction in the amplitude of the total response function, and the period of oscillation is shortened, see Fig.4.
\begin{figure}[h]
\centering 
{\includegraphics[width=.45\textwidth]{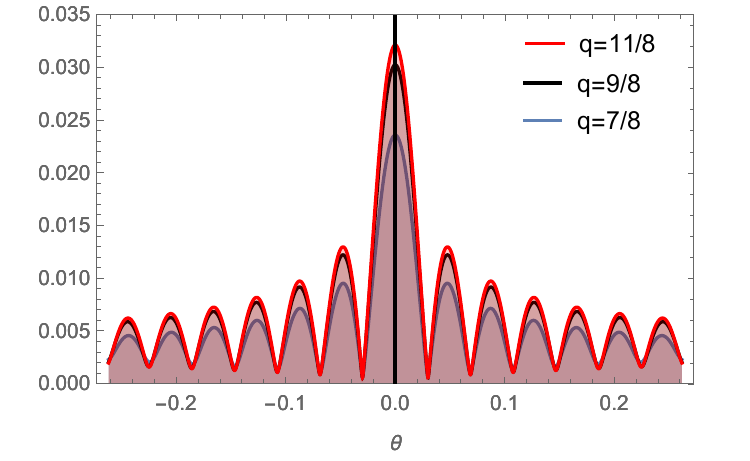}}
\caption{\label{fig3}  The absolute amplitude of the total response function ${\langle \mathcal{O} \rangle} _{\mathcal{J}_O}$ for different  power parameter  $q$ with $Q_{YM}=0.5$, $\xi_+=0.5$  and $\hat{\omega}=80$. }
\end{figure}

\begin{figure}[h]
\centering 
{\includegraphics[width=.45\textwidth]{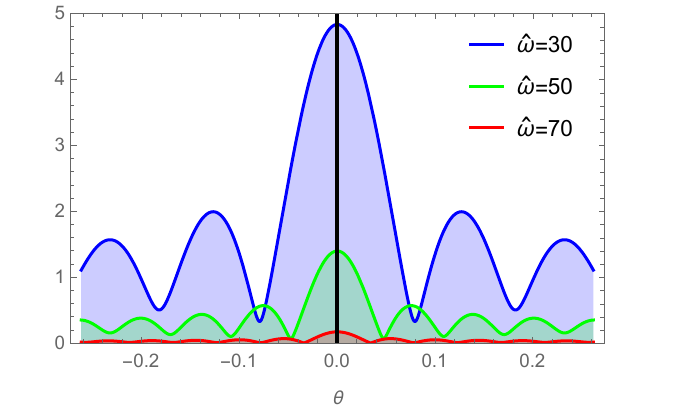}}
\caption{\label{fig4}  The absolute amplitude of the total response function ${\langle \mathcal{O} \rangle} _{\mathcal{J}_O}$  for different   source frequency  $\hat{\omega}$  with $Q_{YM}=0.5$, $\xi_+=0.5$ and $q=7/8$. }
\end{figure}

Subsequently, we also manipulate the temperature $T$ and chemical potential $u$ of the boundary system to elucidate the changes in  behavior of the total response function. The oscillation period of the wave amplitude for different values of $u$ is depicted in Fig.5, with specific values selected for other involved parameters as $\xi_+ = 0.5$, $\hat{\omega} = 80$, and $q = 7/8$. One can observe a decreasing trend in the amplitude of the total response function as the values of $u$ increase. In Fig.6, the amplitude of the total response function exhibits significant temperature dependence when the involved parameters are taken as $Q_{YM}=0.5, \hat{\omega}=80$ and $q=7/8$. At $T=0.626$, the amplitude reaches a higher peak position, followed by a significant reduction at $T=0.513$ and $0.439$, that is,  the amplitude increases with the decrease of the temperature $T$. The behavioral patterns of the relation $|{\langle \mathcal{O} \rangle} _{\mathcal{J}_O}|-T$ and $|{\langle \mathcal{O} \rangle} _{\mathcal{J}_O}|-u$  exhibit certain similarities.

\begin{figure}[h]
\centering 
{\includegraphics[width=.45\textwidth]{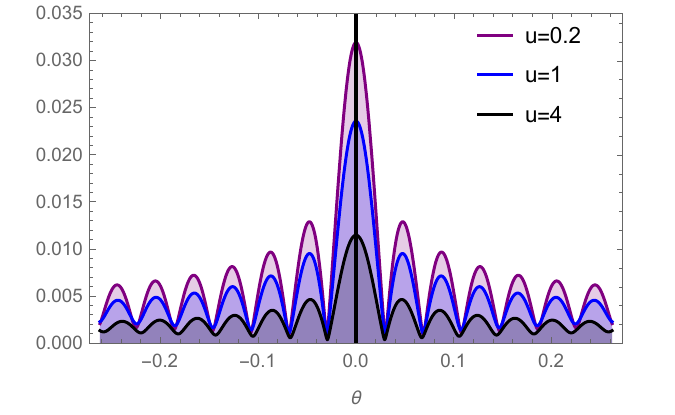}}
\caption{\label{fig5} The absolute amplitude of the total response function ${\langle \mathcal{O} \rangle} _{\mathcal{J}_O}$ for different chemical potential $u$ of the boundary system  with $\xi_+=0.5$, $\hat{\omega}=80$ and $q=7/8$.}
\end{figure}

\begin{figure}[h]
\centering 
{\includegraphics[width=.45\textwidth]{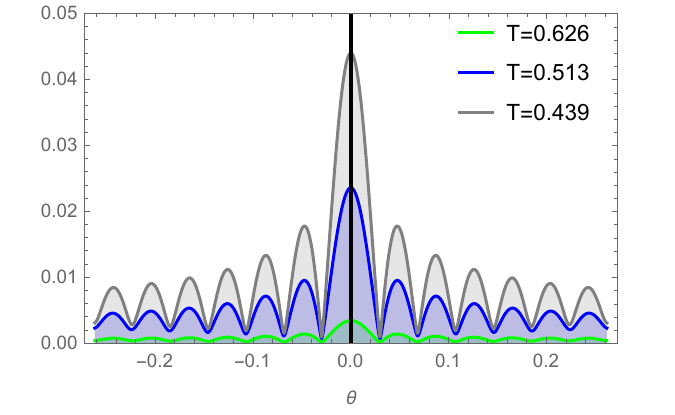}}
\caption{\label{fig6} The absolute amplitude of the total response function ${\langle \mathcal{O} \rangle} _{\mathcal{J}_O}$ for different temperature $T$ of the boundary system  with $Q_{YM}=0.5$, $\hat{\omega}=80$ and $q=7/8$.}
\end{figure}

\section{Holographic Einstein ring of black hole  in Einstein-power-Yang-Mills gravity }
As discussed previously, the response function encompasses a wealth of spacetime information. By utilizing Eq. (\ref{WF}) and the virtual optical system depicted in Fig.2, we can acquire  the profiles of the dual black hole in EPYM gravity projected onto the screen.  The Gaussian source, located at the south pole of the AdS boundary, should be noted. By appropriately selecting specific parameter values, we conduct numerical simulations to accurately depict the  profile images of black holes and  their corresponding brightness curve in Fig.7 to Fig.14.

\begin{figure}[h]
\centering 
\subfigure[${\hat{\omega}}=10$]{\includegraphics[width=.3\textwidth]{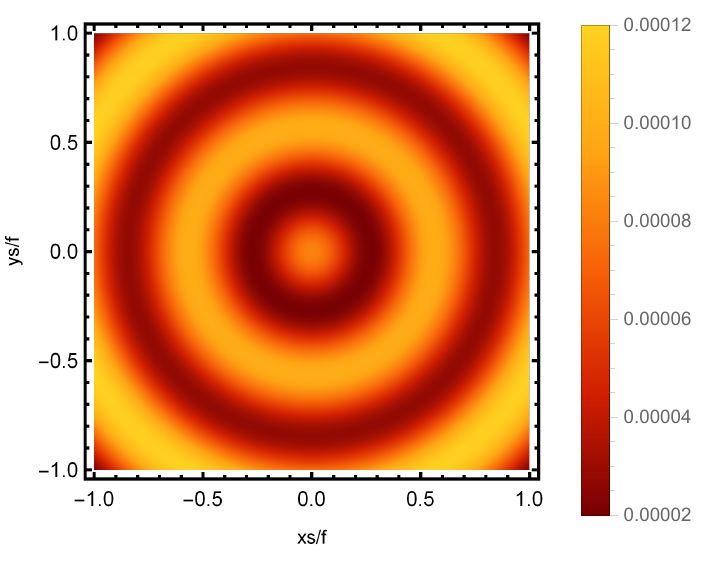}}
\subfigure[${\hat{\omega}}=20$]{\includegraphics[width=.29\textwidth]{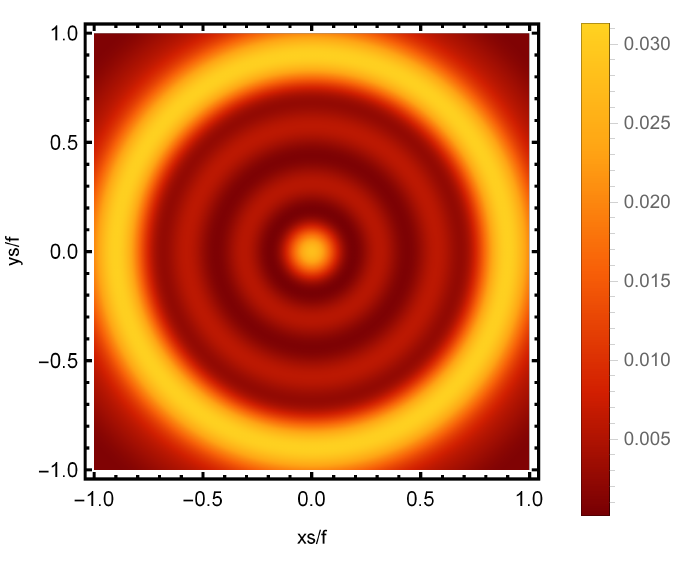}}
\subfigure[${\hat{\omega}}=30$]{\includegraphics[width=.29\textwidth]{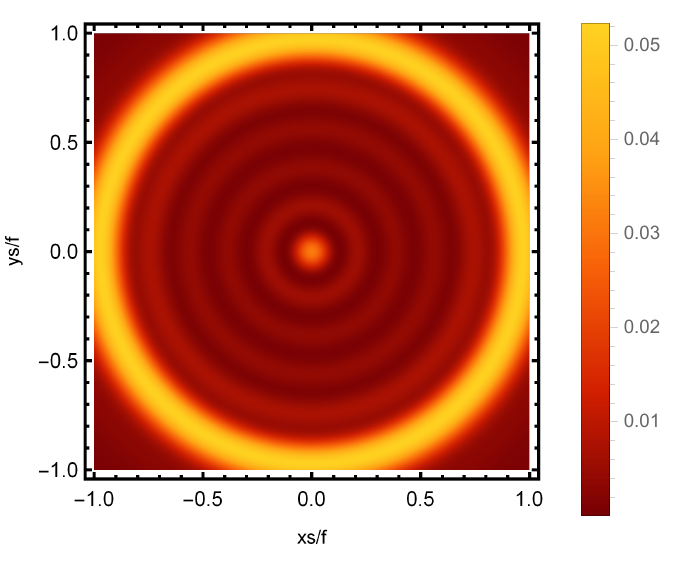}}
\subfigure[${\hat{\omega}}=50$]{\includegraphics[width=.29\textwidth]{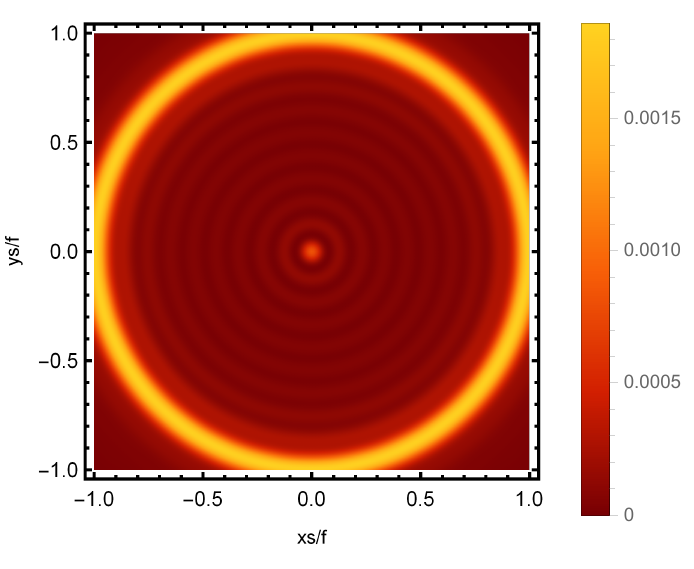}}
\subfigure[${\hat{\omega}}=60$]{\includegraphics[width=.29\textwidth]{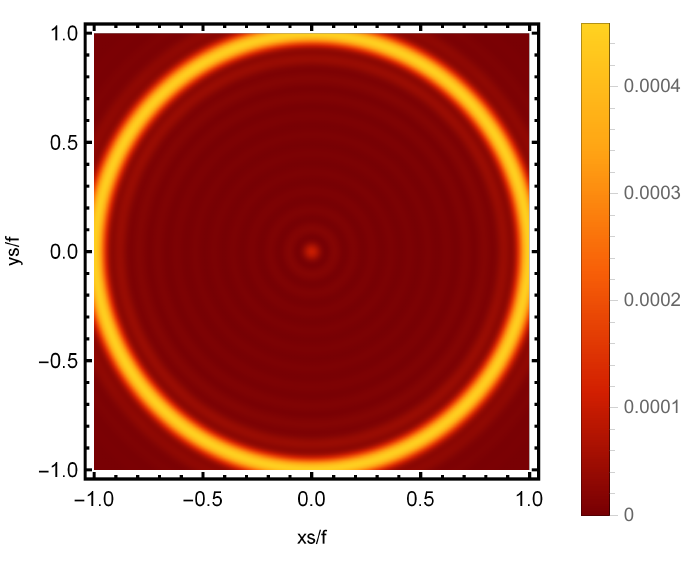}}
\subfigure[${\hat{\omega}}=70$]{\includegraphics[width=.3\textwidth]{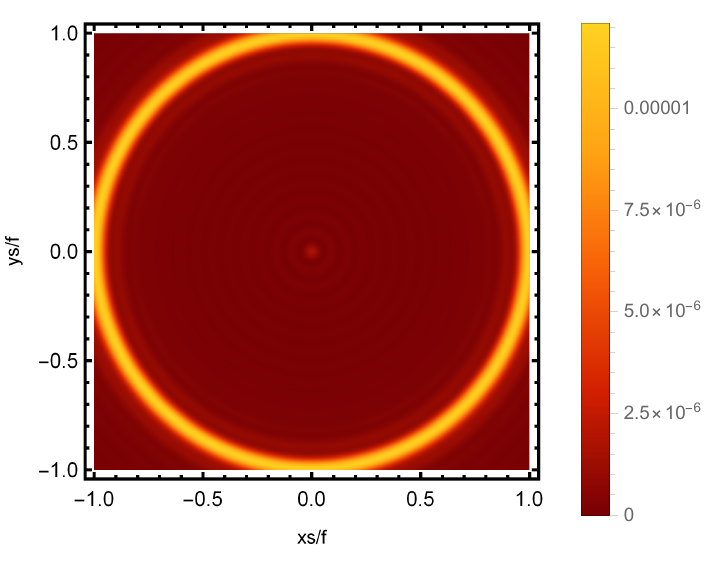}}
\caption{\label{fig7} The images of the lensed response observed at the observation angle $\theta_{obs}=0$  for different $\hat{\omega}$  with $Q_{YM}=0.5$,  $\xi_+=0.5$ and $q=7/8$.}
\end{figure}

In Fig.7, we examine the impact of varying the source frequency $\hat{\omega}$ on the characteristics of the holographic images observed from the north pole of the AdS boundary ($\theta_{obs}=0$), while keeping other relevant parameters fixed at $Q_{YM}=0.5, \xi_+=0.5$ and $q=7/8$. The results demonstrate a distinct  visibility of the striped pattern, revealing a series closed axisymmetric  arrangement of concentric ring. Among them, there is consistently an exceptionally luminous ring, known as the holographic Einstein ring, which corresponds to the photon ring region of the black hole in terms of geometric optics. The width of the ring decreases and becomes sharper as the value of $\hat{\omega}$ increases.

The corresponding brightness curves of the lensed response on the screen are depicted in Fig.8, where the $y$-axis and and $x$-axis represent the intensity and position of  brightness curves, respectively. It is evident that the location of the bright ring in Fig.7 corresponds to where the maximum  peak of the  lensed response brightness curve occurs, and  variations in frequency elicit  changes in the location of the maximum peak. Specifically, the increase in frequency causes the location of the maximum peak to gradually move away from the center of the screen, that is, the radius of the bright ring increase, which is consistent with the result in Fig.7. Furthermore,  the increase in frequency consistently results in a progressively steeper brightness curve, which aligns with the enhanced image sharpness depicted in Fig.7. From this perspective,  the optical appearance of the Einstein image obtained through the geometric optics approximation is highly satisfactory in the high frequency limit.

\begin{figure}[h]
\centering 
\subfigure[${\hat{\omega}}=10$]{\includegraphics[width=.275\textwidth]{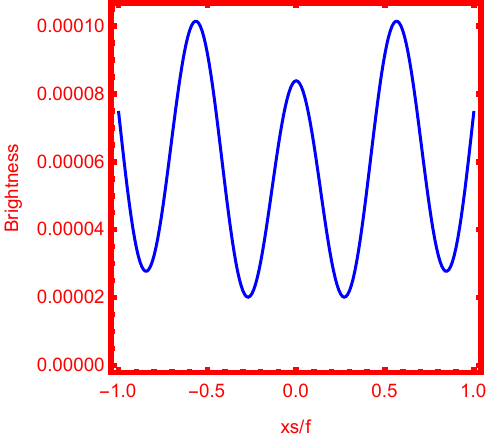}}
\subfigure[${\hat{\omega}}=20$]{\includegraphics[width=.26\textwidth]{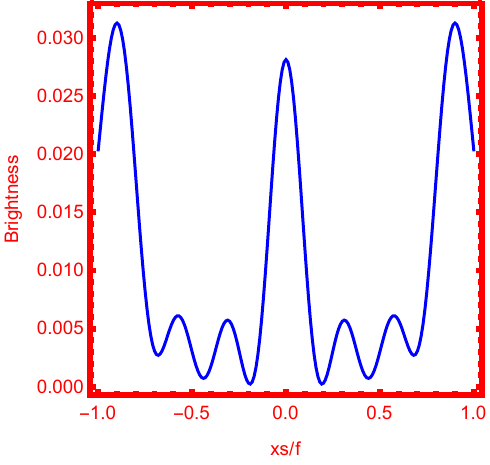}}
\subfigure[${\hat{\omega}}=30$]{\includegraphics[width=.255\textwidth]{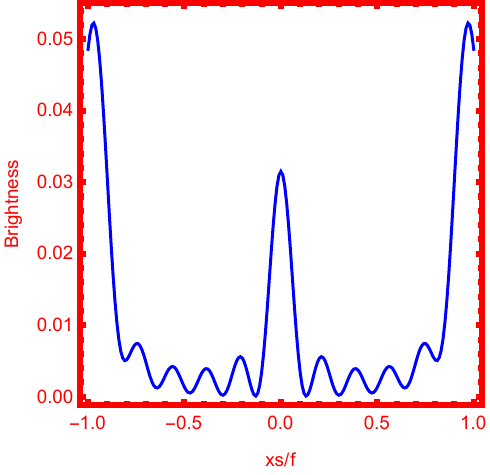}}
\subfigure[${\hat{\omega}}=50$]{\includegraphics[width=.26\textwidth]{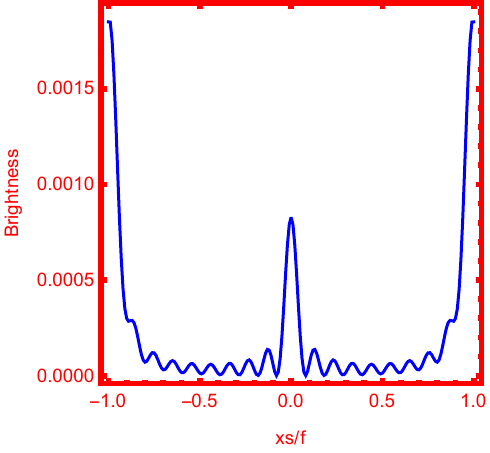}}
\subfigure[${\hat{\omega}}=60$]{\includegraphics[width=.26\textwidth]{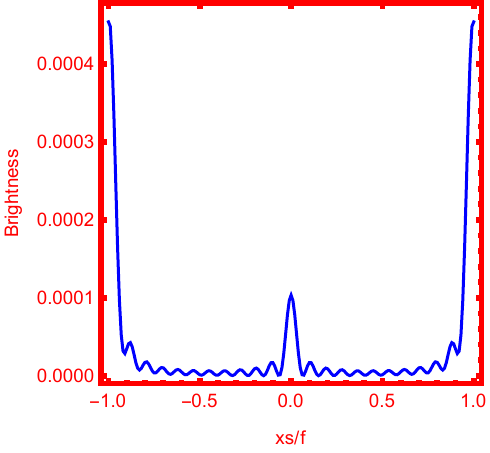}}
\subfigure[${\hat{\omega}}=70$]{\includegraphics[width=.27\textwidth]{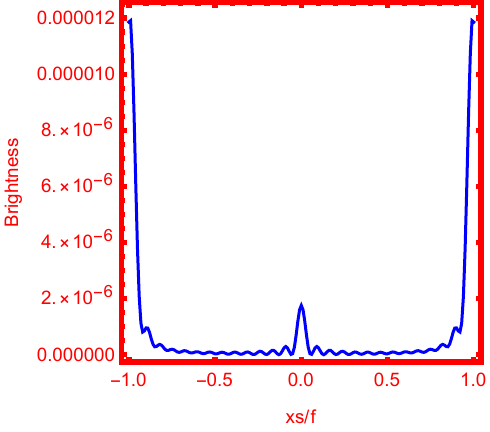}}
\caption{\label{fig8} The brightness of the lensed response on the screen for different $\hat{\omega}$  with $Q_{YM}=0.5$,  $\xi_+=0.5$ and $q=7/8$.}
\end{figure}

We further study the effect of power exponent $q$ on the holographic images of the dual black hole. Given the crucial significance of parameter $q$ in spacetime geometry, we intend to  explore the holographic images from various observation angles. In Fig.9, the corresponding parameter value is set to $Q_{YM}=0.5, \xi_+=0.5, \hat{\omega}=80$, and the value of power exponent $q$ is incrementally augmented. When the observation position is located at the north pole of the AdS boundary $\theta_{obs}=0$, a series of axisymmetric concentric circles rings can still be discernible in the image. As parameter $q$ increases, there is a slight increment in the radius of bright ring, causing its position to shift slightly away from the center of the image, as shown in the left column of Fig.9. Upon repositioning the observation angle to  $\theta_{obs}=\pi/4$, the axisymmetric bright ring dissipates and gives way to a prominent bright arc positioned on the left side of the image, as illustrated in the middle column of Fig.9. Notably, a small and less bright arc also appears on the right side of the image, exhibiting a decrease in size and luminosity as $q$ increases.  When $q=11/8$, the bright arc on the right side of the image essentially  disappears. As the observation angle at $\theta_{obs}=\pi/2$, a solitary bright spot emerges solely on the left side of the image, and as parameter $q$ increases, the position of this bright spot progressively shifts away from the image center. Under identical parameter settings, the brightness distribution curve of the holographic image at $\theta_{obs}=0$ can be obtained, as illustrated in Fig.10. It can be observed that as the parameter $q$ increases, there is a slight shift in the position of the maximum peak of the brightness curve, gradually moving away from the center. In other words, an increase in the power exponent $q$ will result in an expansion of the radius of the holographic ring.

\begin{figure}[h]
\centering 
\subfigure[$q=7/8$,$\theta_{obs}=0$]{\includegraphics[width=.25\textwidth]{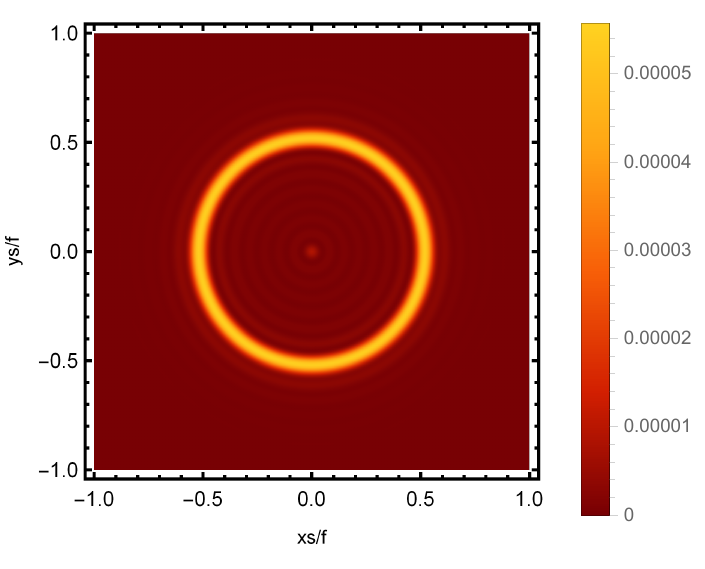}}
\subfigure[$q=7/8$,$\theta_{obs}=\pi/4$]{\includegraphics[width=.25\textwidth]{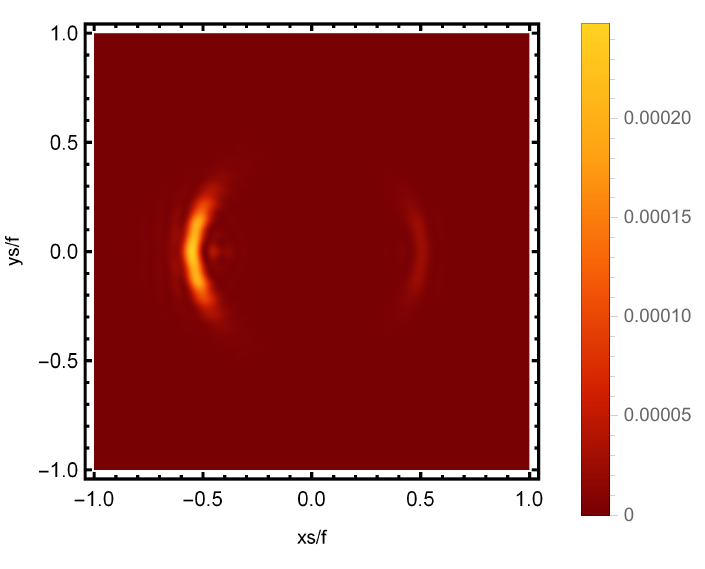}}
\subfigure[$q=7/8$,$\theta_{obs}=\pi/2$]{\includegraphics[width=.25\textwidth]{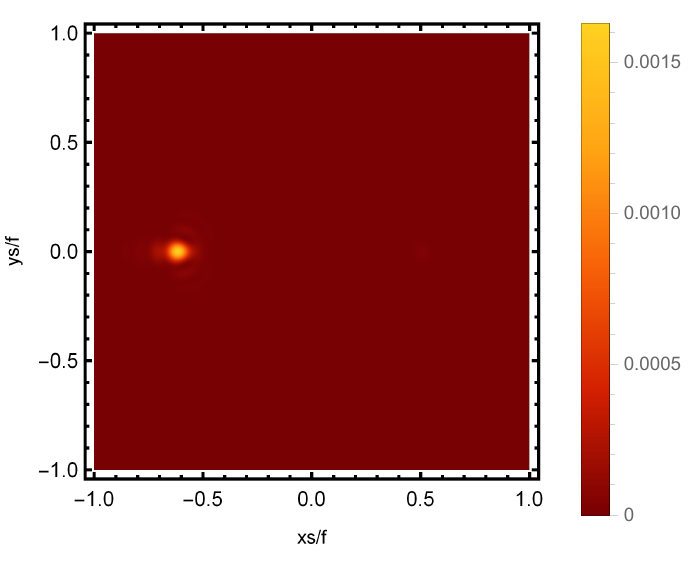}}
\subfigure[$q=9/8$,$\theta_{obs}=0$]{\includegraphics[width=.25\textwidth]{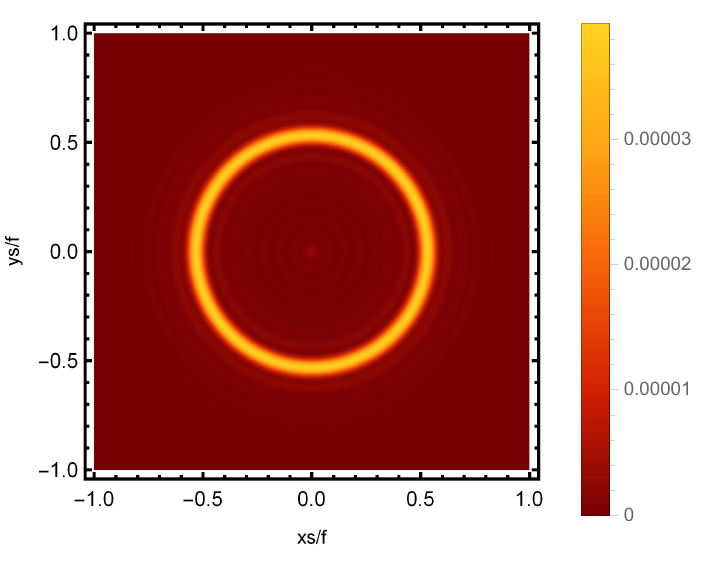}}
\subfigure[$q=9/8$,$\theta_{obs}=\pi/4$]{\includegraphics[width=.25\textwidth]{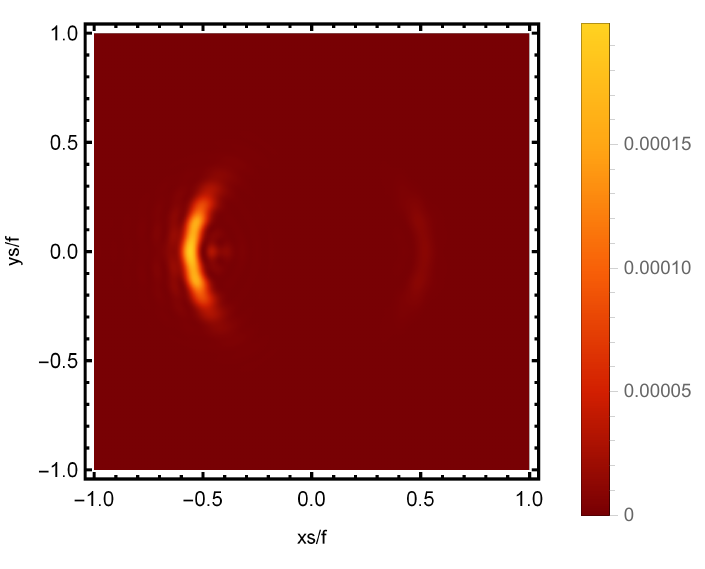}}
\subfigure[$q=9/8$,$\theta_{obs}=\pi/2$]{\includegraphics[width=.25\textwidth]{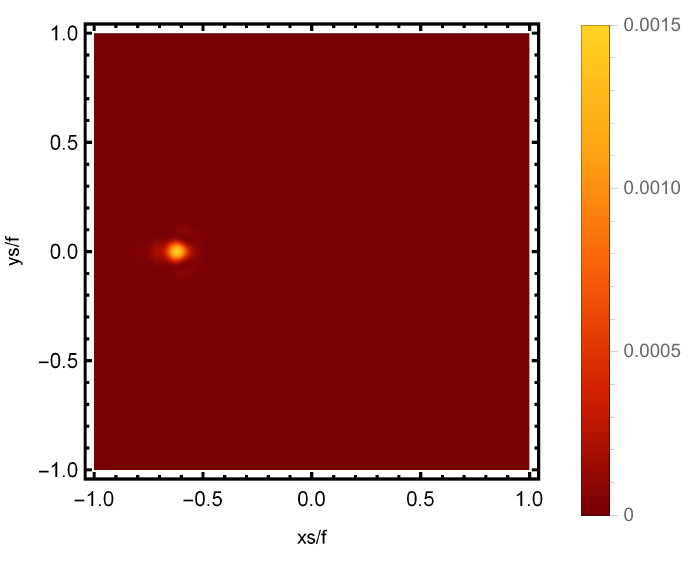}}
\subfigure[$q=11/8$,$\theta_{obs}=0$]{\includegraphics[width=.25\textwidth]{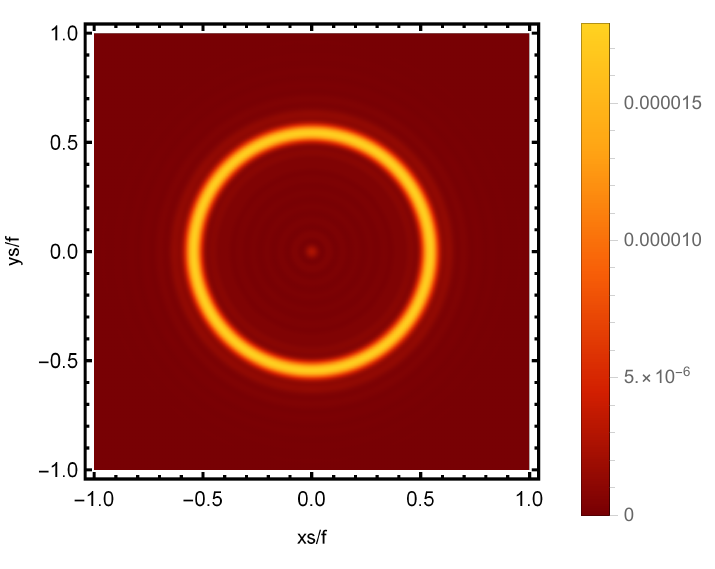}}
\subfigure[$q=11/8$,$\theta_{obs}=\pi/4$]{\includegraphics[width=.25\textwidth]{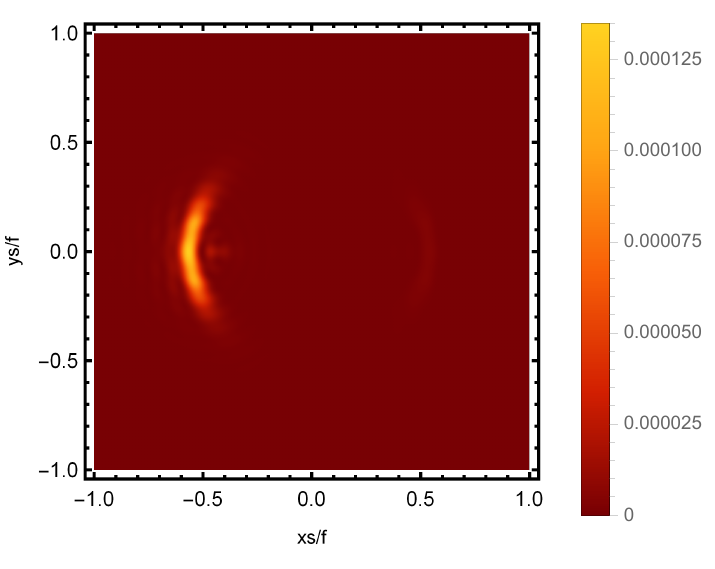}}
\subfigure[$q=11/8$,$\theta_{obs}=\pi/2$]{\includegraphics[width=.25\textwidth]{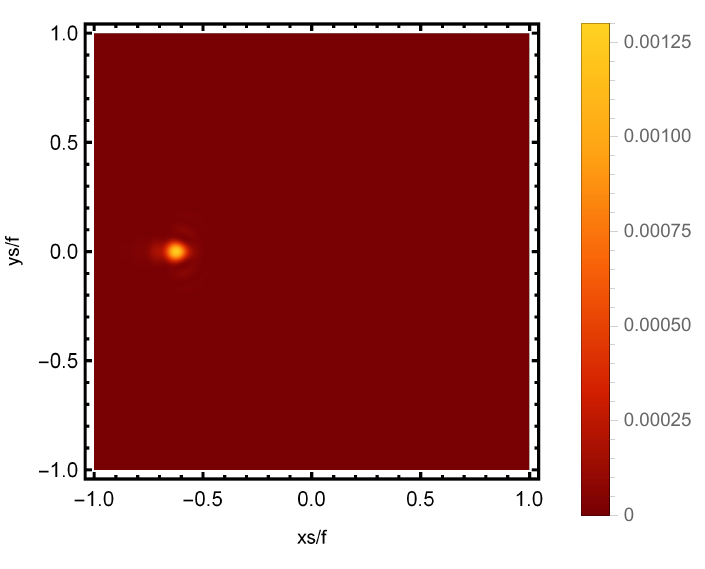}}
\caption{\label{fig9} The images of the lensed response observed at various observation angles for different $q$ with $Q_{YM}=0.1$,  $\xi_+=5$ and $\hat{\omega}=80$.}
\end{figure}
\begin{figure}[h]
\centering 
\subfigure[$q=7/8$]{\includegraphics[width=.25\textwidth]{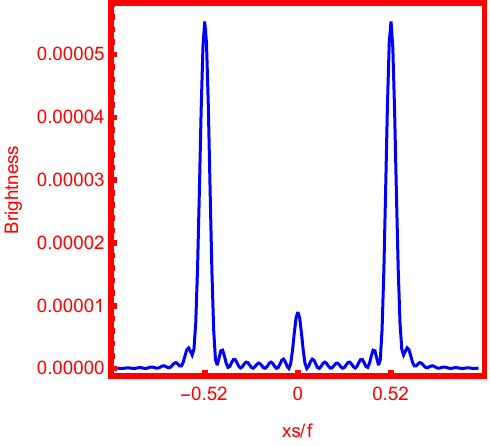}}
\subfigure[$q=9/8$]{\includegraphics[width=.25\textwidth]{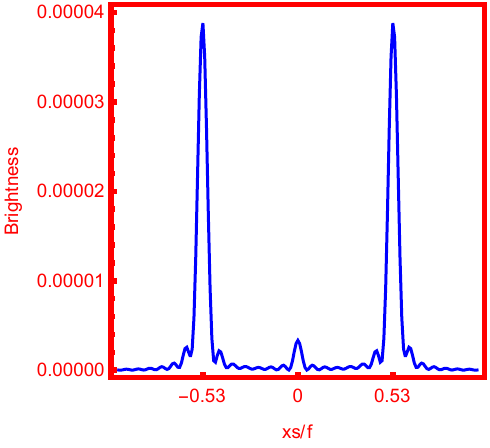}}
\subfigure[$q=11/8$]{\includegraphics[width=.25\textwidth]{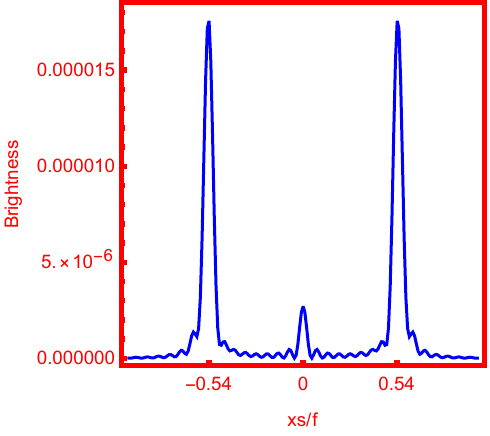}}
\caption{\label{fig10} The brightness of the lensed response on the screen for different $q$ with $\hat{\omega}=80$, $Q_{YM}=0.1$ and  $\xi_+=5$.}
\end{figure}

Subsequently, we are going to investigate the effect of the chemical potential $u$  on the images of the lensed response.  These images  are observed under fixed values of relevant parameters such as  $\theta_{obs} = 0, q = 7/8$ and $\hat{\omega} = 80$, are shown in Fig.11. As the value of $u$ increases from $u = 0.3$ to $u = 7$, the radius of the brightest ring exhibits a barely noticeable decrease, but abruptly diminishes like a precipice when $u=0.9$. The same phenomenon is observed in the brightness curve of the lens response, which is shown in Fig.12. It can be seen that as the chemical potential $u$ increases, the peak position of the curve gradually approaches the central position, but this trend is not sufficiently pronounced, except for a notable displacement at $u=0.9$. The resulting Einstein ring radius is obviously dependent on the chemical potential $u$, especially the case of higher chemical potential, which is different from the result\cite{Liu:2022cev}. This discrepancy arises due to the chemical potential of the boundary system is modified under the background of EPYM gravity, which naturally has a significant effect on the radius of the ring.

\begin{figure}[t]
\centering 
\subfigure[$u=0.3$]{\includegraphics[width=.225\textwidth]{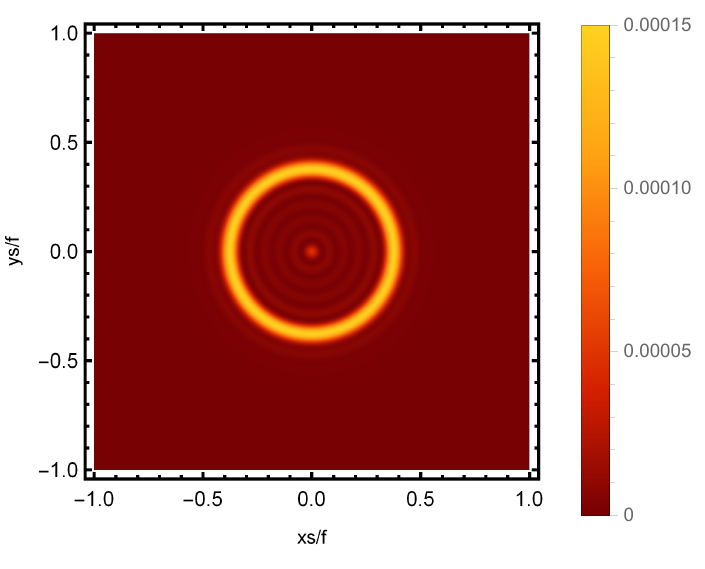}}
\subfigure[$u=0.5$]{\includegraphics[width=.225\textwidth]{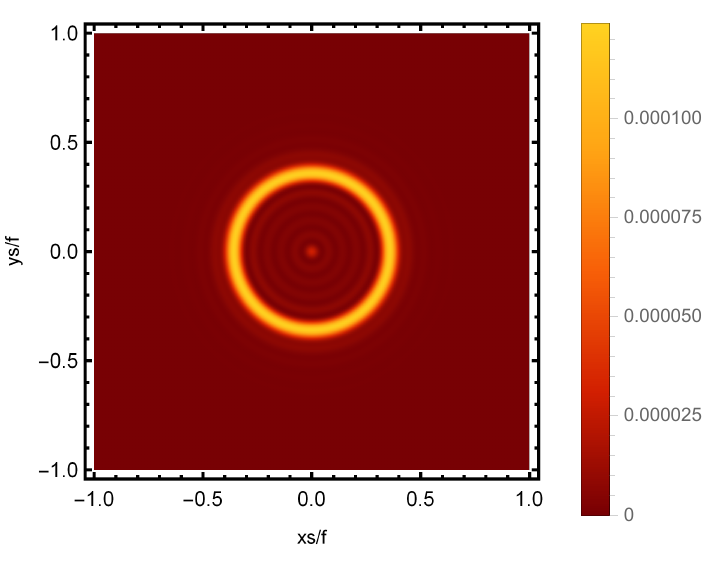}}
\subfigure[$u=0.7$]{\includegraphics[width=.225\textwidth]{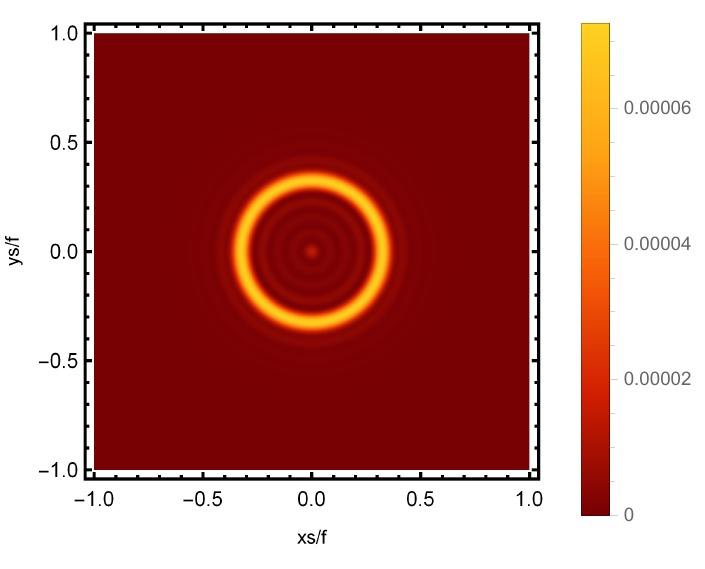}}
\subfigure[$u=0.9$]{\includegraphics[width=.225\textwidth]{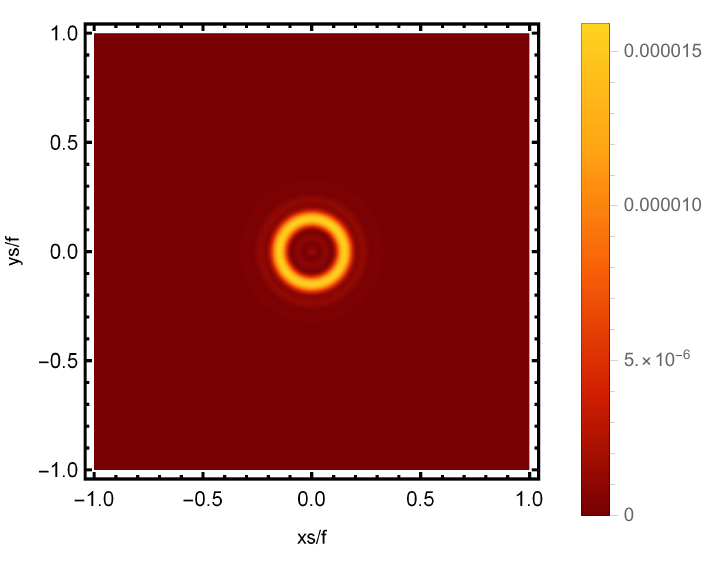}}
\caption{\label{fig11} The images of the lensed response observed at the observation angle $\theta_{obs}=0$ for different chemical potential  $u$ with a fixed temperature  $T=0.6$. Here $\hat{\omega}=80$ and $q=7/8$.}
\end{figure}
\begin{figure}[t]
\centering 
\subfigure[$u=0.3$]{\includegraphics[width=.225\textwidth]{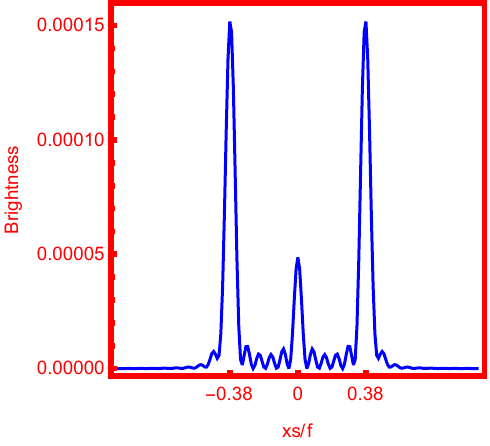}}
\subfigure[$u=0.5$]{\includegraphics[width=.225\textwidth]{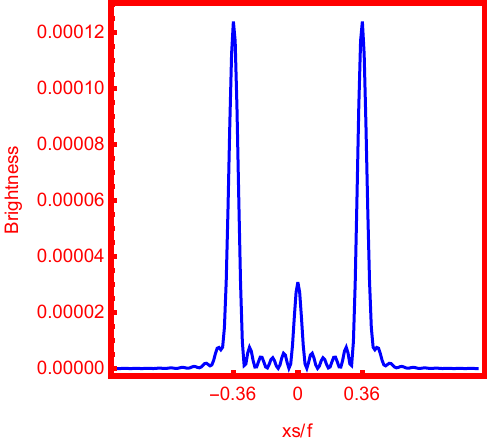}}
\subfigure[$u=0.7$]{\includegraphics[width=.225\textwidth]{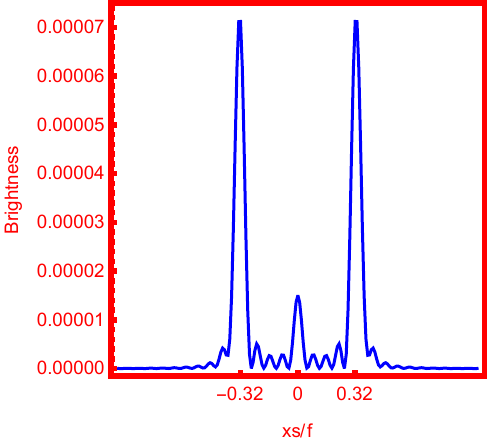}}
\subfigure[$u=0.9$]{\includegraphics[width=.225\textwidth]{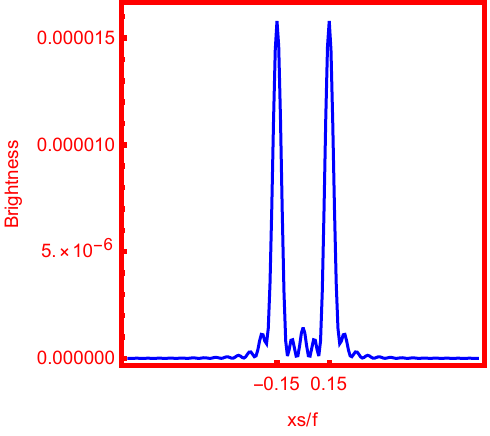}}
\caption{\label{fig12} The brightness of the lensed response on the screen for different chemical potential  $u$ with a fixed temperature  $T=0.6$. Here $\hat{\omega}=80$ and  $q=7/8$.}
\end{figure}

The variation of the image obtained at the north pole with different temperatures is plotted in Fig. 13, while keeping the other parameters fixed at $u = 1$, $\hat{\omega} = 80$, and $q = {7}/{8}$.  When the temperature is $T=0.229$, $0.277$, $0.472$, and $0.730$, the corresponding values of the charge parameter $Q_{YM}$ in the system are $Q=0.2$, $0.5$, $0.8$, and $1.1$, respectively. As we can see, the radius of the formed bright ring has a visible increase as the temperature rises within the range of $T=0.229$ to $T=0.277$. In the temperature range of $T=0.277$ to $T=0.730$, although there is a certain increasing trend of the radius of ring, it is not  significant. From the  brightness curve  of the lensed response in Fig.14, it reveals a trend of increasing distance between the maximum peak of the curve and the center of the screen, aligning precisely with the  alteration in bright ring radius.  The temperature dependence of the radius of the Einstein ring is evident. Specifically, at low temperatures, the radius of the ring significantly increases with an increase in temperature, which aligns with the findings in \cite{Liu:2022cev}.

\begin{figure}[h]
\centering 
\subfigure[$T=0.229$]{\includegraphics[width=.225\textwidth]{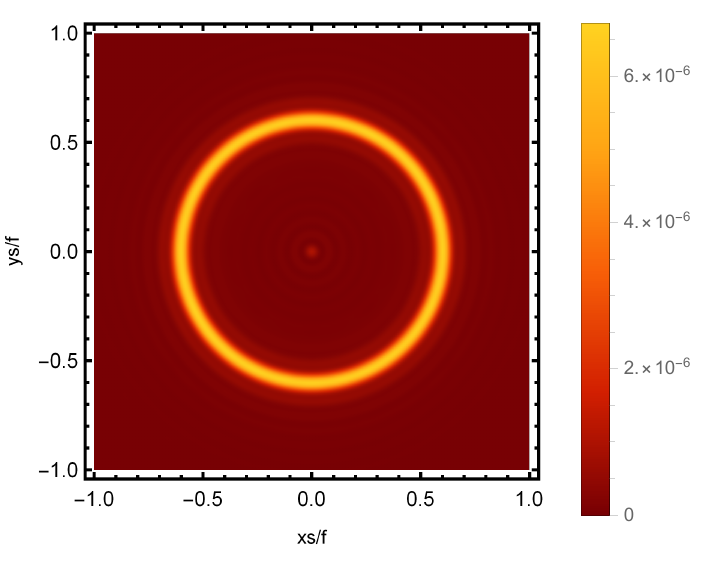}}
\subfigure[$T=0.277$]{\includegraphics[width=.225\textwidth]{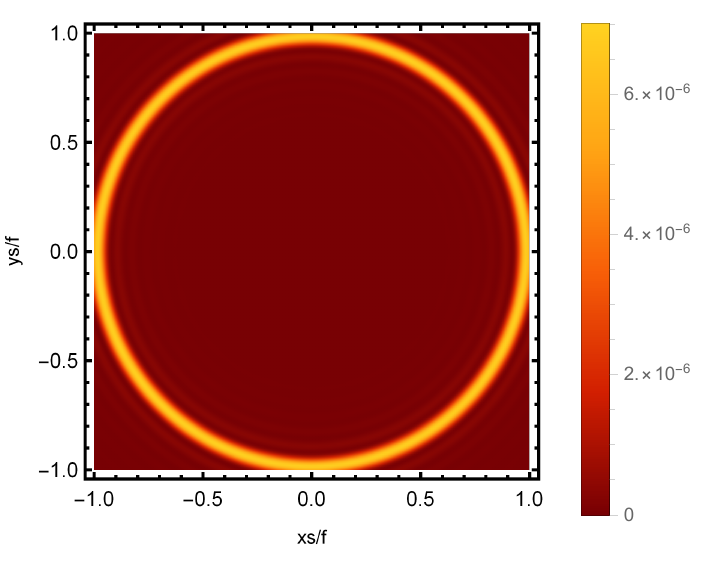}}
\subfigure[$T=0.472$]{\includegraphics[width=.225\textwidth]{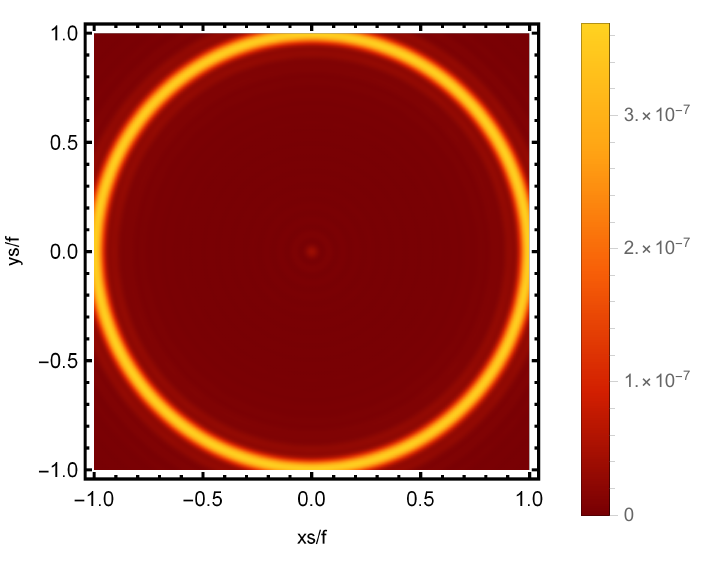}}
\subfigure[$T=0.730$]{\includegraphics[width=.225\textwidth]{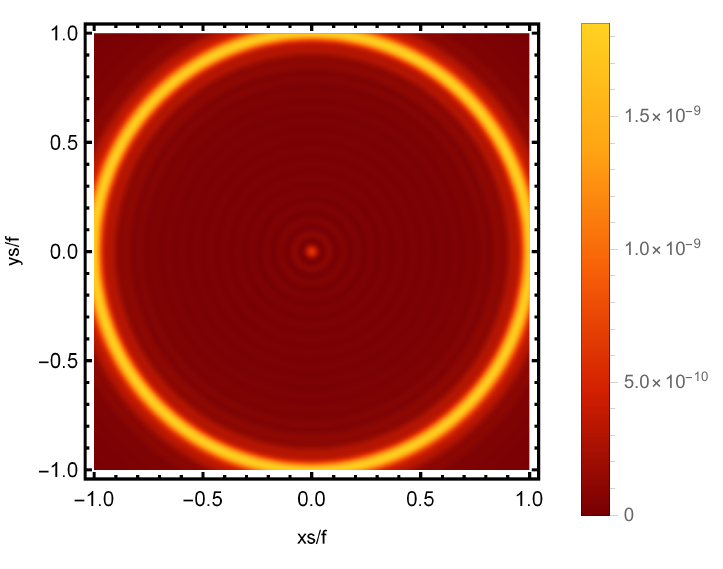}}
\caption{\label{fig13} The images of the lensed response observed at the observation angle $\theta_{obs}=0$  for different $T$ with a fixed $u=1$. Here, $\hat{\omega}=80$, $q=7/8$ and from (a) to (d), the charges correspond  to $Q_{YM}=0.2, 0.5, 0.8, 1.1$ respectively.}
\end{figure}
\begin{figure}[h]
\centering 
\subfigure[$T=0.229$]{\includegraphics[width=.225\textwidth]{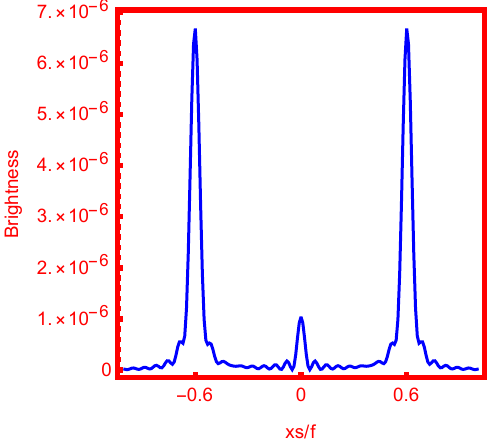}}
\subfigure[$T=0.277$]{\includegraphics[width=.225\textwidth]{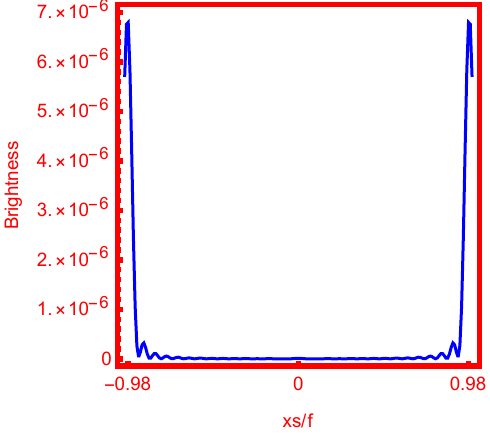}}
\subfigure[$T=0.472$]{\includegraphics[width=.225\textwidth]{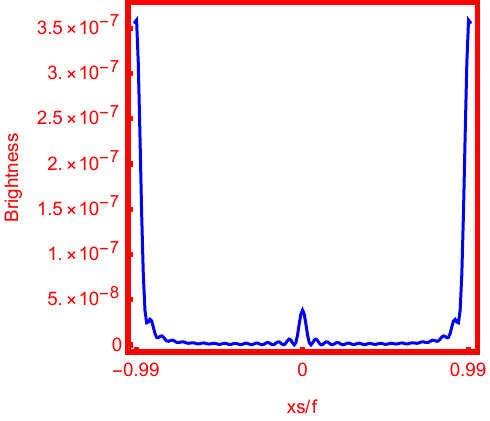}}
\subfigure[$T=0.730$]{\includegraphics[width=.225\textwidth]{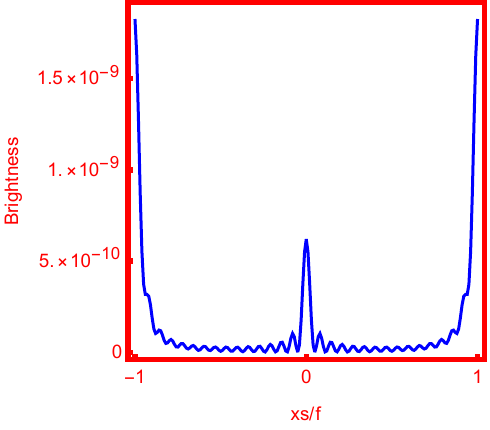}}
\caption{\label{fig14} The brightness of the lensed response on the screen for different $T$ with a fixed $u=1$. Here, $\hat{\omega}=80$, $q=7/8$ and from (a) to (d), the charges correspond  to $Q_{YM}=0.2, 0.5, 0.8, 1.1$ respectively.}
\end{figure}

Following\cite{Liu:2022cev}, the relationship between the linear response function and the source can be characterized by the retarded Green function $\mathcal{G} (t, \theta, \varphi; t', \theta', \varphi')$, that is
\begin{align}
{\langle \mathcal{O} \rangle} _{\mathcal{J}_{\mathcal{O}}}=\int^{2 \pi}_0 d \varphi ' \int^{\pi}_0 d \theta' \sin \theta ' \mathcal{G}(t, \theta, \varphi; t', \theta', \varphi') \mathcal{J}_{\mathcal{O}}(t', \theta', \varphi')=\sum _{\zeta=0}^{\infty } e^{-i \hat{\omega}  t} \mathcal{G}_{\zeta 0}(\hat{\omega}) C_{{\zeta 0}} Y_{\zeta 0}(\theta). \label{Green}
\end{align}
The retarded Green function $\mathcal{G} $ for a weakly interacting QFT with the finite temperature or chemical potential can be read as
\begin{align}
\mathcal{G}_{\varsigma m}(\hat{\omega})=\frac{1}{\hat \omega^2-\varsigma(\varsigma+1)-m_T^2}. \label{Green2}
\end{align}
Here, $m_T$ is the thermal mass and can be expressed as
\begin{align}
m_T^2=m^2+\frac{\rho}{2} h(T,u), \label{Tmass}
\end{align}
where the  correction term $h(T,u)$ is
\begin{align}
h=\sum_\varsigma \frac{1}{\hat{\omega}_\varsigma} \left[\frac{1}{\mathit{e}^{\alpha(\hat{\omega}_\varsigma+u)}-1}+\frac{1}{\mathit{e}^{\alpha(\hat{\omega}_\varsigma-u)}-1}\right].\label{Correction}
\end{align}
The parameter $\alpha$ represents  the inverse of temperature, given by $\alpha=\frac{1}{T}$, and $\hat{\omega}_\varsigma=\sqrt{m^2+\varsigma(\varsigma+1)}$. The variation tendency of $h$ allows us to qualitatively determine the response or ring radius, thus we present the  $h$  typical dependence on the temperature $T$ and chemical potential $u$ in Fig.15.

\begin{figure}[t]
\centering 
\subfigure[]{\includegraphics[width=0.3\textwidth]{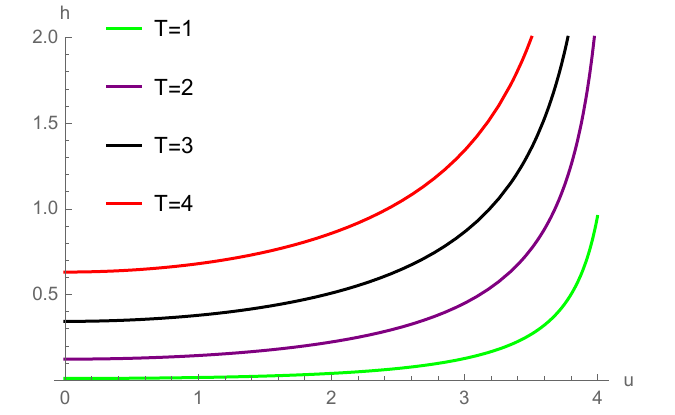}}
\subfigure[]{\includegraphics[width=0.3\textwidth]{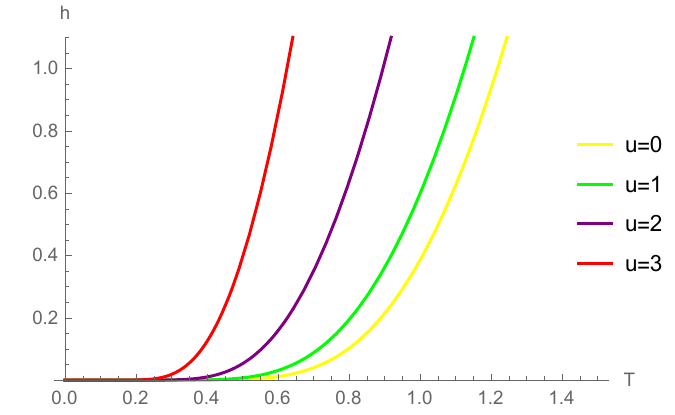}}
\caption{\label{fig15}  The dependence of $h(t,u)$ on the temperature and the chemical potential
with $m^2=-2$.}
\end{figure}
As depicted in Fig.15 (a), for a given chemical potential, an increase in the temperature $T$ leads to an increase in the value of $h$, and the radius of the corresponding ring becomes larger. The value of $h$ increases over a very small interval in the temperature range of $0<T<1$, indicating that the change in the radius of the ring within this temperature range will not be significant. These results are consistent with those shown in Fig.13. From Fig.15 (b), the variation of $h$ exhibits negligible for the chemical potential $0<u<1$ in a low temperature $T$, which is  consistent with that obtained by Fig.13. However, at a fixed temperature $T$, the value of $h$ increases  with larger $u$, indicating that the radius of the ring should correspondingly increase as well. This does not match the results shown in Fig.11.

\subsection{Comparison between holographic and optical results}

Regarding the shadow image of a black hole, it is established that the luminous ring depicted in the image corresponds  to the location of the photon ring of the black hole. In order to delve deeper into the physical significance of the holographic Einstein ring, we will analyze the correlation between the bright ring in the  image and the photon ring surrounding a black hole from a geometric optics perspective. We can employ the Eikonal approximation as it transforms the Klein-Gordon equation into the form of  Hamilton-Jacobi equation, denoted by
\begin{align}
g^{\mu\nu}(\partial_\mu \mathcal{I} -\mathit{e}{A}_t)(\partial_\nu \mathcal{I}-\mathit{e}{A}_t)-\mathbb{M}^2=0, \label{Hamilton}
\end{align}
where $\mathcal{I}$ stands for the action. In the system (\ref{metric1}), the above equation can be written as
\begin{align}
-N(r)^{-1}(\partial_t \mathcal{I}-\mathit{e}{A}_t)^2+N(r)\partial_r \mathcal{I}^2+\frac{1}{r^2} \partial^2_{\theta}+\frac{1}{r^2 \text{sin}^2 \theta} \partial^2_{\varphi}-\mathbb{M}^2=0. \label{metricH}
\end{align}
The spherically symmetric  structures in Eq. (\ref{metric1}) allow us to always position the  photon orbits on the equatorial plane, simplifying the model without sacrificing generality, i.e.,  $\theta=\frac{\pi}{2}$ and $\dot\theta=0$. In this spacetime,  there exist two killing vector $\omega$ and $L$, enabling the expression of the action as
\begin{align}
\mathcal{I}(t,r,\varphi)=-\omega t+L\varphi+\int N(r)^{-1}\sqrt{\mathcal{R}}dr,\label{Action}
\end{align}
and
\begin{align}
\mathcal{R}=(\omega-\mathit{e}{A}_t)^2-N(r)(\frac{L^2}{r^2}-2).\label{Action2}
\end{align}
In which
\begin{align}
\omega=-\frac{\partial \mathcal{I}}{\partial t}, \quad
L=\frac{\partial \mathcal{I}}{\partial \varphi},\quad
\frac{\sqrt{\mathcal{R}}}{N(r)}=\frac{\partial \mathcal{I}}{\partial r}.\label{MFA}
\end{align}
The term of $\omega$ and $L$ denoted as the conserved energy and angular momentum, respectively. The path of photon propagation will be described by the aforementioned conditions. Then, the ingoing angle $\theta_{in}$ \cite{Liu:2022cev} with the normal vector of boundary $\mathfrak{n}^\gamma =\frac{ \partial}{\partial r^{\gamma}}$ is defined as
\begin{align}
\cos \theta_{in}=\frac{g_{ij}\mathfrak{u}^i \mathfrak{n}^j}{|\mathfrak{u}||\mathfrak{n}|}\mid_{r=\infty}=\sqrt{\frac{\dot{r}^2/N}{\dot{r}^2/N+L/r^2}}\mid_{r=\infty}.\label{IngoingC}
\end{align}
 One can also get
\begin{align}
\sin \theta_{in}^2=1-\cos \theta_{in}^2=\frac{L^2 N /r^2}{\dot{r}^2+L^2 N/r^2} \mid_{r=\infty}=\frac{L^2}{\omega^2}. \label{IngoingS}
\end{align}
Hence, the ingoing angle $\theta_{in}$ of photon orbit from the boundary  is thus determined as
\begin{align}
\sin \theta_{in}=\frac{L}{\omega}. \label{IngoingS2}
\end{align}
The photons in close proximity to the photon ring will orbit around the black hole one or more times, resulting in cumulative luminosity within this region.  The photons that can be located in the vicinity of the photon ring also applies to Eq. (\ref{IngoingS2}), and the corresponding diagram is shown in Fig. 16.
\begin{figure}[t]
\centering 
{\includegraphics[width=0.45\textwidth]{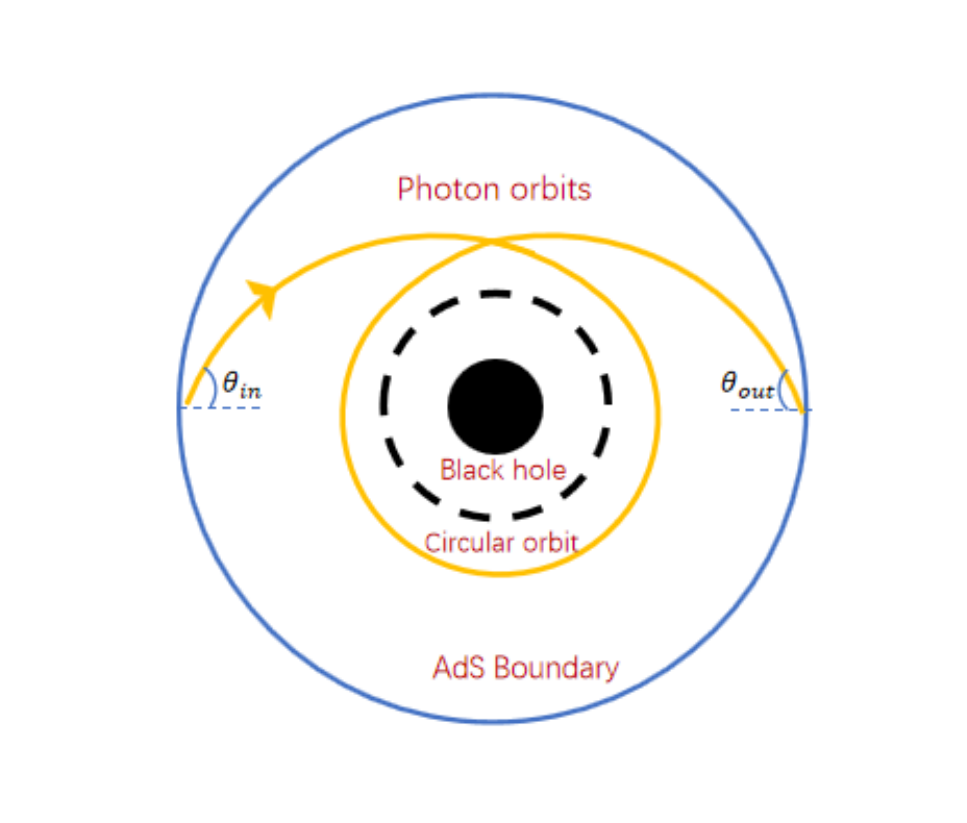}}
\caption{\label{fig16}  The ingoing angle and outgoing angle  of the photon at the photon ring.}
\end{figure}
Here, the angular momentum of these photons is denoted as $L_S$, and their light trajectory satisfies the following conditions
\begin{align}
\mathcal{R}=0, \quad
 \frac{d \mathcal{R}}{d r}=0. \label{PH}
\end{align}
From the perspective of geometric optics, the angle $\theta_{in}$ represents the angular displacement of the incident ray's image from the zenith when an observer positioned on the AdS boundary gazes upwards into the AdS bulk. In a scenario where the entry and exit points of photon trajectories on the AdS boundary align precisely with the center point of the black hole, the radius of the  ring image observed by an observer corresponds to the incidence angle $\theta_{in}$ due to axial symmetry. And the Einstein ring is formed on the screen, as shown in Fig.17, with the ring radius as
\begin{align}
\sin \theta_{R}=\frac{r_R}{f}.\label{radius3}
\end{align}
\begin{figure}[t]
\centering 
{\includegraphics[width=0.8\textwidth]{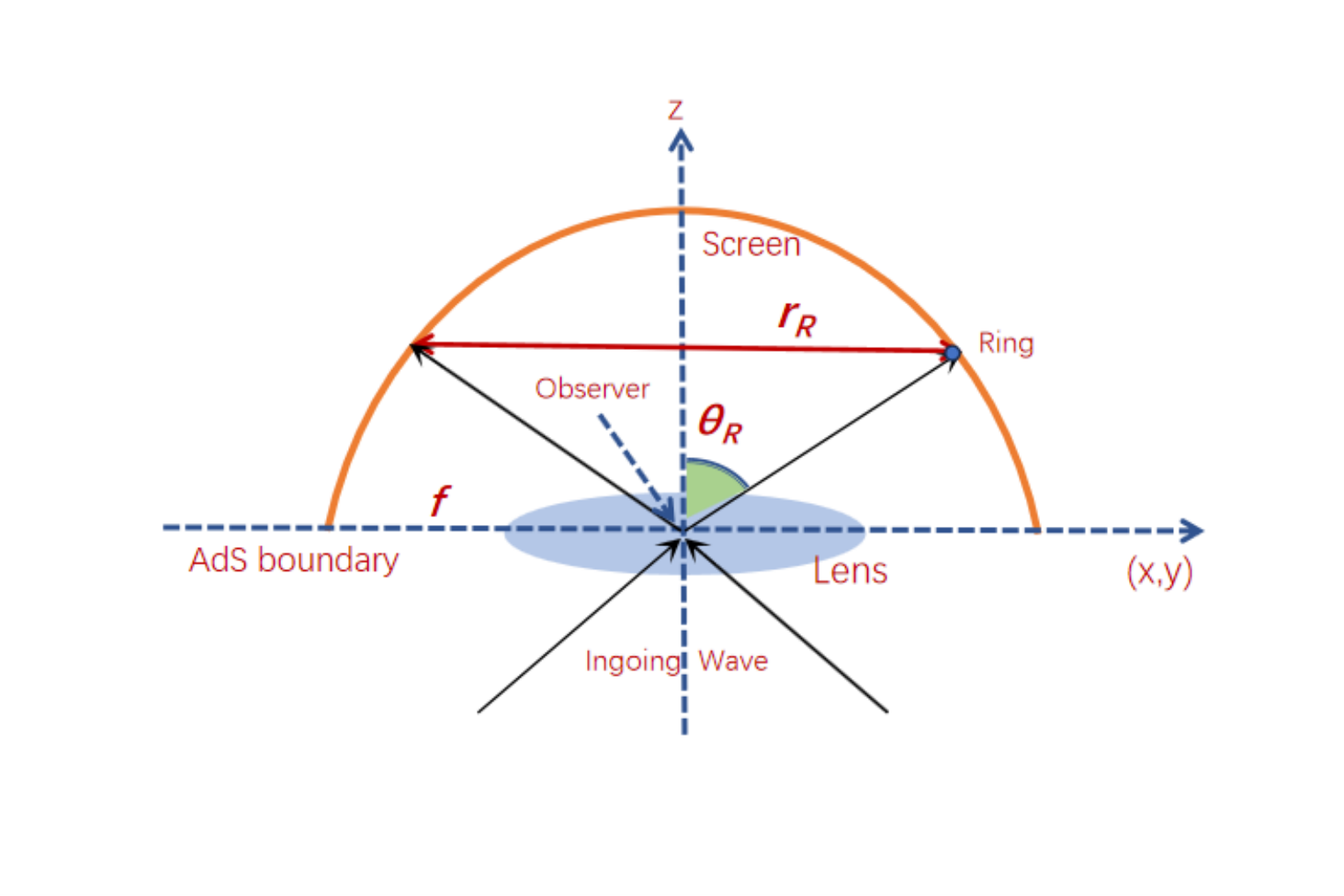}}
\caption{\label{fig17}  The relation between the $\theta_R $ and $r_R$.}
\end{figure}
According to \cite{Hashimoto:2019jmw,Hashimoto:2018okj}, one can obtain $ \sin \theta_{R}= \sin \theta_{in}$ for a sufficiently large $\varsigma$, which means that
\begin{align}
\frac{r_R}{f}=\frac{L_S}{\omega}, \label{AG}
\end{align}
and the focal length $f$ of the lens here determines the degree of magnification. Therefore, the radius  of holographic ring is in accordance with the size of the photon ring. In essence, the incident angle of the photon and the angle of the photon ring both refer to the angle of the photon ring that the observer can observe, and they should be basically equivalent. Simultaneously, we will validate this relationship through numerical. In Fig.18 and 19, the radius of the Einstein ring is depicted for different values of the parameter $q$, where $r_R/ f$  represents the ring radius as a function of temperature and chemical potential. It can be observed that the red data points consistently cluster around the blue curve, exhibiting values within a three percent margin of each other. The Einstein ring angle obtained in the holographic analysis is consistent with the ingoing angle derived from geometric optics, thus demonstrating a high level of agreement between the two approaches.
\begin{figure}[t]
\centering 
\subfigure[$q=7/8$]{\includegraphics[width=0.3\textwidth]{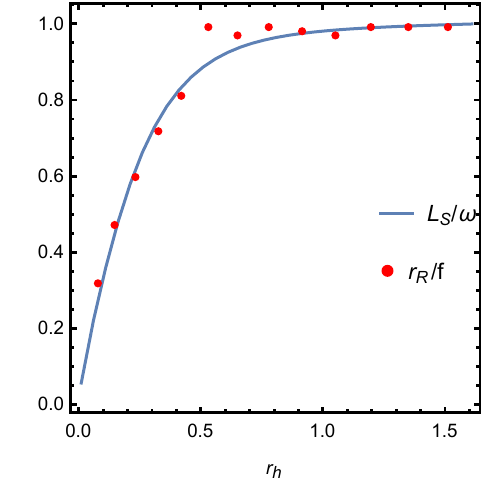}}
\subfigure[$q=9/8$]{\includegraphics[width=0.3\textwidth]{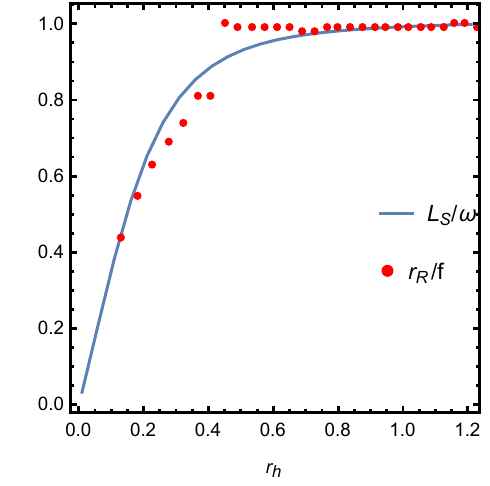}}
\subfigure[$q=11/8$]{\includegraphics[width=0.3\textwidth]{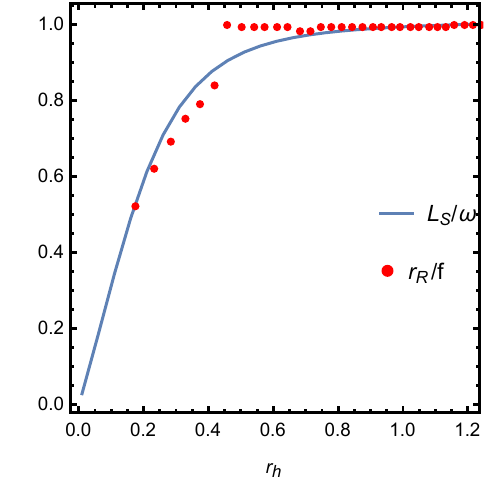}}
\caption{\label{fig18}  Comparison between  the results obtained by wave optics and geometric optics for different $q$ with a fixed chemical potential $u=1$.}
\end{figure}

\begin{figure}[t]
\centering 
\subfigure[$q=7/8$]{\includegraphics[width=0.3\textwidth]{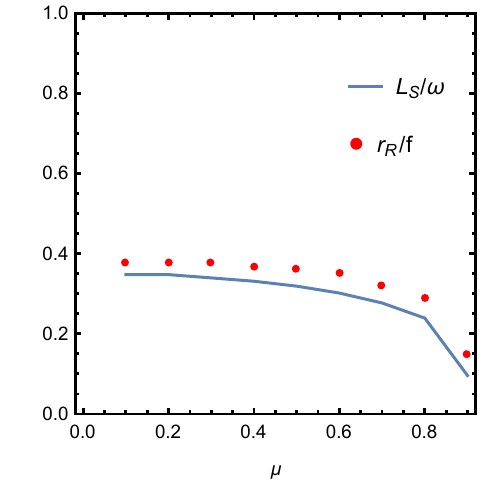}}
\subfigure[$q=9/8$]{\includegraphics[width=0.3\textwidth]{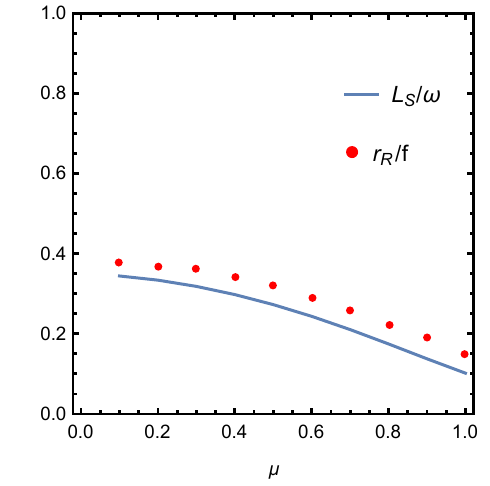}}
\subfigure[$q=11/8$]{\includegraphics[width=0.3\textwidth]{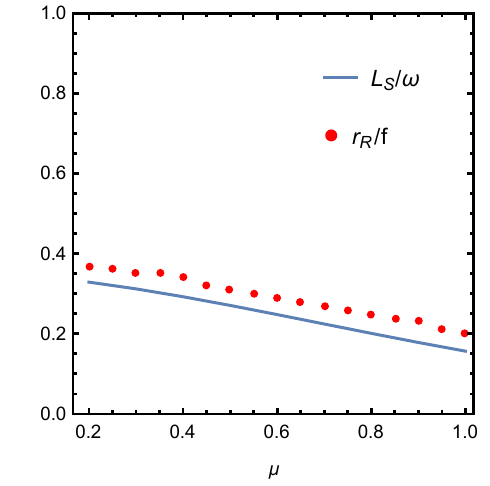}}
\caption{\label{fig19} Comparison between  the results obtained by wave optics and geometric optics for different $q$ with a fixed temperature $T=0.6$.}
\end{figure}

\section{Conclusions and discussions}
\label{conclusion}
Gravitational lensing, a prevalent optical phenomenon in the universe, serves as a paramount research tool and pivotal means within astrophysics. It assumes a momentous role in cosmology encompassing dark matter, dark energy, large-scale gravity, and exoplanet detection.  As one of the most vivid demonstrations of gravitational lensing, Einstein rings offer an abundance of valuable insights into the intricate fabric of spacetime. From this perspective, we employ wave optics to investigate the holographic  image of an AdS black hole within EPYM gravity framework. The AdS boundary is equipped with an oscillating Gaussian source, and the scalar wave emitted from this source propagates through the  bulk spacetime, resulting in a corresponding lensed response on the opposite side of the boundary. The diffraction pattern of the total response function ${\langle \mathcal{O} \rangle} _{\mathcal{J}_{\mathcal{O}}}$ is always present during scalar wave propagation through the black hole, as revealed by extracting the local response function at the boundary. The characteristics of the absolute amplitude of the total response function, for various values of the involved parameters, are depicted in Fig.\ref{fig3}-Fig.\ref{fig6}. These figures demonstrate how wave oscillations change in response to variations in state parameters. From Fig.\ref{fig3}, it can be seen that with the increase of the  power exponent $q$, the amplitude peak of ${\langle \mathcal{O} \rangle} _{\mathcal{J}_{\mathcal{O}}}$ shows a steady increasing trend. However, as the frequency of the wave source increases, the amplitude peak of ${\langle \mathcal{O} \rangle} _{\mathcal{J}_{\mathcal{O}}}$ diminishes, and the time period of the scalar wave shortens, see Fig.\ref{fig4}. Furthermore, the impacts of the finite chemical potential $u$  and horizon temperature $T$ on the amplitude of the total response function are illustrated in Fig.\ref{fig5} and  Fig.\ref{fig6}. The elevation in temperature $T$ and chemical potential $u$ will progressively diminish the corresponding amplitude of the response function. However, the reduction is more pronounced with variations in temperature $T$.

Although the diffraction pattern of the response function is analyzed, it does not intuitively provide specific information about the black hole. By utilizing a virtual optical system consisting of  a convex lens and a spherical screen ( Note that this optical system is in neither the bulk spacetime nor the boundary), we are able to capture the holographic image of an  AdS black hole in EPYM gravity. In all cases, irrespective of the specific parameters involved, when the light source is situated at the south pole, an observer located at the north pole ($\theta_{obs}=0$) will invariably perceive an image manifesting as a luminous, axisymmetric ring, commonly referred to as the holographic Einstein ring. Certainly, variations in the relevant parameters exhibit distinct impacts on specific attributes of the ring, such as its size and brightness. In Fig.\ref{fig7}, it is observed that as the value of $\hat{\omega}$ increases, the width of the ring becomes progressively narrower and more sharply defined, while the radius of the ring exhibits a decreasing trend. At the same time, the brightness curve of the response function is also provided, where the  maximum  peak on the curve corresponds to the position of the bright ring, see Fig.\ref{fig8}. The results demonstrate that as frequency increases, the position of the maximum peak progressively shifts away from the center of the screen. In other words, the radius of the bright ring expands, which is consistent with the findings presented in Fig.\ref{fig7}. Interestingly, in low frequency bands, an increase in frequency leads to an increase in observed brightness, while in high frequency bands, an increase in frequency leads to a decrease in observed brightness. In Fig.\ref{fig11}, we analyzed the influence of the chemical potential $u$ on the obtained images of AdS black hole in EPYM gravity. The results indicate that the radius of the ring decreases as the chemical potential $u$ of the boundary system increases. However, this decrease is relatively insignificant at low chemical potentials ($0.3<u<0.5$) and becomes more pronounced as the chemical potential reaches higher values ($0.7<u<0.9$), which  is different from the effect of chemical potential on ring radius in \cite{Liu:2022cev}. Due to the chemical potential of the boundary system is modified under the background of EPYM gravity, the impact of changes in chemical potential on the radius of the resulting ring is emphasized.  From Fig.\ref{fig12}, it can be observed that the brightness curve of the  lensed response on the screen, corresponding to different chemical potentials $u$, shows a gradual decrease in the intensity of the observable ring brightness as $u$ increases.
The impact of boundary system temperature $T$ variation on the ring  is contrary to that of chemical potential, that is, the radius of the ring increases when the temperature increases, especially at low temperature. The influence of temperature on the ring radius is consistent with the results in\cite{Liu:2022cev}.

Considering the dependence of spacetime on the the power exponent $q$, we also investigate the effect of the change of $q$ on the Einstein ring at different observation angles ($\theta_{obs}=0, \pi/4, \pi/2$), as shown in Fig.\ref{fig9}.  As depicted in the leftmost column of Fig.\ref{fig9}, when the observation angle is $\theta_{obs}=0$, one can find that the image of the black hole appears as a bright ring with a series of concentric stripes, which correspond to the diffraction behavior of the total response function. Then, the radius of the bright ring in the image is observed to increase with an increasing parameter $q$, and the same conclusion can be obtained from the brightness curve in Fig.\ref{fig10}. When the observation angle is $\theta_{obs}=\pi/4$,  as illustrated in the middle column of Fig.\ref{fig9}, the ring structure in the image becomes disrupted, leaving  a bright arc visible on the left side of the image, while the right side shows a smaller and less intense arc. The bright arc on the left side remains unaffected as the parameter $q$ increases, but the bright arc on the right side gradually diminishes. In the rightmost column of Fig.\ref{fig9}, the observation angle is $\theta_{obs}=\pi/2$, the bright arc disappears and only one bright spot is located on the left side of the image. With the increase of parameter $q$, the position of the spot tends to move away from the center of the image. Moreover, the observable brightness of the resulting image will decrease as  parameter $q$ increases, regardless of any changes in the observation angles.

We end  with comparing the results between the bright ring obtained within the holographic framework and the position of the photon ring of  black hole obtained through geometric optics. In Fig.\ref{fig18} and \ref{fig19},
The radii of the photon ring (depicted by the blue curve) and the Einstein ring (represented by discrete red dots) are plotted as functions of varying parameter $q$ values. As anticipated,  the ingoing angles of the photon ring consistently align numerically with those of the Einstein ring across varying temperatures $T$ and chemical potentials $u$. Hence, a correlation can be established between the positions of the photon ring and the holographic Einstein ring, thereby affirming the reliability of wave optics within the holographic framework. In addition, holographic images can serve as a crucial tool for discerning the geometric characteristics of different black holes, and also providing  specific insights into the phenomenological consequences of black hole dynamics, especially when considering fixed wave sources and optical systems. We aspire for these discoveries to furnish a more vivid and intuitive understanding of Einstein rings and their associated phenomena, thereby facilitating future tabletop experiments.

\vspace{10pt}

\noindent {\bf Acknowledgments}

\noindent
This work is supported by the National Natural Science Foundation of China (Grants No. 11875095), Innovation and Development Joint Foundation of Chongqing Natural Science Foundation (Grant No. CSTB2022NSCQ-LZX0021),
Basic Research Project of Science and Technology Committee of Chongqing (Grant No. CSTB2023NSCQ-MSX0324) and the Natural Science Foundation of Chongqing (CSTB2023NSCQ-MSX0594).



\end{document}